\documentclass[twocol]{ametsocV6.1_arXiv}

\usepackage{multirow}
\usepackage{amsmath}

\title{Improving prediction of heavy rainfall in the Mediterranean with Neural Networks using both observation and Numerical Weather Prediction data}

\authors{Killian Pujol,\aff{a}\correspondingauthor{Killian Pujol, killian.pujol--nicolas@univ-tlse3.fr} 
Roberta Baggio,\aff{b}
Dominique Lambert,\aff{a}
Jean-François Muzy,\aff{b} 
Jean-Baptiste Filippi,\aff{b} 
Florian Pantillon\aff{a} 
}

\affiliation{\aff{a}{Laboratoire d’Aérologie (LAERO), Université de Toulouse, CNRS, IRD, UT, Toulouse, France} \\
\aff{b}{Laboratoire Sciences Pour L’Environnement (SPE), Université de Corse, CNRS, Corte, France}}

\abstract{Forecasting Heavy Precipitation Events (HPE) in the Mediterranean is crucial but challenging due to the complexity of the processes involved. In this context, Artificial Intelligence methods have recently proven to be competitive with state-of-the-art Numerical Weather Prediction (NWP). This work focuses on improving the prediction of the occurrence of HPE over periods from 1 h to 24 h based on Neural Network (NN) models. The proposed method uses both ground-station observations and data from Météo France’s Arome and Arpege NWP models, on two regions with oceanic and Mediterranean climates for the period 2016-2018. The verification metric is the Peirce Skill Score. Results show that the NN model using only observations or NWP data performs better for shorter and longer rainfall accumulation period respectively. In contrast, a hybrid method combining both observations and NWP data offers the best performance and remains stable with the rainfall accumulation period. The hybrid method also improves the performance in predicting increasingly intense rainfall, from the 5\% to the 0.1\% rarest events. The choice of the loss function is found to be an important aspect of this work, where only balanced loss functions provide results insensitive to rare event frequency. Finally, the hybrid method is particularly well suited for the prediction of HPE in the Mediterranean climate, especially during the fall season, period during which most HPE occur.}

\begin{document}

\maketitle

%
%
%
%
%
%

%
%
%
%
%
%
%
\section{Introduction}

The Mediterranean basin is characterized by hot, dry summers and mild, wet winters. It presents a significant meteorological variability due to interactions between tropical and polar air masses, together with a complex topography characterized by high mountain ranges. These peculiarities can lead to extreme weather events such as intense rainfall and storms. Heavy Precipitation Events (HPEs) in particular are a cause of major concern and can generate flash floods whose impact on economy and society often results in human fatalities. 
The list of catastrophic cases is long, from historic floods caused by an Aiguat in Catalonia in 1940 \citep{Aiguat} to cyclone Daniel that brought heavy rainfall in Greece and Lybia \citep{egusphere-2024-2809}
or the more recent flooding in the Valencia region of Spain \citep{Valence}.
As the Mediterranean region is densely populated, understanding and forecasting the meteorological dynamics of these events is crucial, even more so with a changing climate \citep{IPCC}. 

Knowledge about the processes responsible for the occurrence of HPEs in the Mediterranean has improved through the years, especially with the research program HyMeX conducted between 2010 and 2020 \citep{HyMeX2}. After summer, the warm Mediterranean sea fills the lower atmosphere with moisture and heat. The air near the sea surface becomes conditionally unstable, i.e., it will ascend vertically if it is triggered by a lifting mechanism. The most common mechanism is lifting by orography: if a flow of moist and conditionally unstable air faces a mountain range, the air may start to ascend and trigger a thunderstorm. This scenario is common in the Mediterranean, as it is surrounded by steep orography, e.g., the Pyrenees between Spain and France, the Atlas mountains throughout Morocco, Algeria and Tunisia, the Taurus mountains in Turkey, as well as mountainous islands like Corsica or Crete. This broad orography next to a warm Mediterranean sea are important ingredients in the formation of rainfall.

Moreover, the triggering of HPE is partly determined from the synoptic-scale context that guides the moist airflow towards the orography. For instance, \citet{Nuissier2011} and \citet{Ricard2012} showed the synoptic environment before and during HPE in the French Mediterranean: a mid-tropospheric trough deepening over the Bay of Biscay, whose orientation and intensity influence the orientation of moisture fluxes, and thus the location of HPE. This is due to the influence of upper-tropospheric instabilities on surface disturbances, as shown by earlier studies in the Mediterranean \citep{Argence2006,Argence2009}. Depending on the synoptic-scale context, the moisture origin of HPE can differ and the quantity of precipitation as well. \citet{Duffourg2011} further showed that both local and remote sources of moisture can feed HPE in cyclonic conditions, with moisture coming from the Atlantic Ocean or tropical Africa, and the added moisture influence greatly the rainfall intensity.

Although the necessary conditions to trigger HPE in the Mediterranean are well known, challenges persist to fully understand their location, intensity and duration. \citet{Bresson2009} studied different lifting effects and how they impact the precise time and location of convective rainfall. Though orographic lifting is the main mechanism, vertical updraft can also be triggered by low-level convergence and cold pools---cold air formed during a thunderstorm by rain evaporation---and the location of the events can vary depending on the lifting mechanism. The authors also emphasized that the location is dependent of several features, such as the speed and instability of the upstream flow.
The duration and stationarity of HPE is another important aspect: the longer the event lasts, the more precipitation might accumulate on a specific location. \citet{Ducrocq2008} studied three cases of stationary HPE and showed that the stationarity depends on the ability of the atmosphere and topography, through blocking and convergence processes, to focus moist and unstable flow at a certain location for a long time. 

Forecasting HPE traditionally relies on Numerical Weather Prediction (NWP) models. These models simulate the evolution of the atmosphere by solving the equations of fluid mechanics and thermodynamics. The initial state is formed by gathering multiple observations from several sources, such as weather stations, radiosondes, precipitation radars and satellites, through a data assimilation process to determine the complete state of the atmosphere. NWP models fall into two types: global NWP models that simulate the atmospheric circulation over the entire Earth, with horizontal resolution starting from 10~km depending on the model, and local NWP model that simulate the weather in smaller, country-sized areas, with horizontal resolution down to about 1~km. At such resolution, the representation of the atmosphere becomes fine enough to solve deep convection without using parametrization schemes, and allows to better represent orography. Overall, high-resolution NWP have been a great advance for HPE forecasts in the Mediterranean, along with improved parametrization schemes and data assimilation \citep[see][for a review of outcomes of the HyMeX program]{Khodayar2021}. 

However, despite these advances for HPE in the Mediterranean, NWP models still face limitations in predicting their precise characteristics. A first source of uncertainties is the initial conditions, as the observations are sparse, especially over sea, and can contain flaws. The uncertainties can then be propagated during simulations \citep{Scheffknecht2016}. Moreover, even if NWP model resolution has increased, small-scale processes such as turbulence cannot yet be explicitly resolved. In addition, other small scale processes, such as air-sea interaction or clouds microphysics \citep{Hally2014}, are not yet well represented in NWP model physics. Finally, even though available computational capacity has increased over the years, the computational cost increases exponentially with NWP models resolution \citep{bauer2015}, and the improvement of NWP models resolution remains a technological challenge.  

Recently, the development of artificial intelligence (AI) has had a major impact on the world of science. Among AI approaches, Machine Learning and Deep Learning are research fields that focus on developing statistical algorithms that are able to learn a specific task from data without explicit instructions for performing that task \citep{Murphy2022}. In particular, Neural Networks (NN) are a type of Deep Learning algorithm that can learn complex spatiotemporal patterns directly from data \citep{Goodfellow2016}. The use of these algorithms in science correlates with the increase in collected data in many areas of scientific research.
Weather data has been widely collected for many years through observations and forecasts, making weather forecasting suitable for the use of AI. As a result, in the last years, many data-driven meteorological models developed for weather forecasting have proven very effective \citep[e.g.][for global weather forecasting]{Lam2023,Pangu}.

Among the earliest approaches exploring the use of Deep Learning for rainfall prediction, many focused on nowcasting, that is, the forecasting of weather evolution in the very short term, typically up to 6 h in the future. Most of these approaches use only observational data as input, especially radar data (e.g. \citet{shi2015,agrawal2019,ayzel2020}). By combining convolutional and recurrent NN for images and time series of observed radar echoes, respectively, 
these studies outperform NWP predictions and classical time extrapolation methods, especially for very short lead times of less than 1 h. However, for longer lead times and for increasingly intense rainfall, the model performances drop significantly. Alternatively, a dense network of weather stations can also be used as a reliable data source for rainfall prediction \citep{WMO2017}. For instance, \citet{pirone2023} proposed a method to forecast the precipitation rate up to 3~h at a given station by using a fully connected NN with present and past observations from the same station as well as neighboring stations as input. This method performs well for the forecast of light to moderate rain, but faces a drop of performance with increasing lead times and for heavy rainfall coming from convective motions. 
As weather stations also record meteorological variables other than precipitation, similar approaches that consider neighboring stations to leverage spatiotemporal dependencies have been used to predict temperature or wind speed from observed fields (e.g. \citet{Baile2023} \citet{Baggio2024}). 
Some studies combine different sources of observational data. For instance, the latest developments of the Google DeepMind MetNet model, MetNet-2 \citep{espeholt2022} and  MetNet-3\citep{andrychowicz2023}, take as input observational data from radar and weather stations as well as satellite and data assimilation products from a high-resolution NWP model. In NWP, data assimilation consists in combining past forecast outputs with observational data in order to obtain the initial conditions to be used for next model initialization, a process which helps in correcting model errors while filtering out erroneous or inconsistent observational values. The prediction of precipitation issued by MetNet-3 drastically outperforms the NWP for short lead-times up to about 6~h, but this performance gap progressively reduces for longer lead times. Moreover, the model performances tends to be lower for higher precipitation rates.

As NN are very effective in learning underlying nonlinear relations and  are capable to combine different type of data, they are also natural candidates for the post-processing of NWP forecasts. These are known to be affected by systematic biases and errors, in particular in terms of  variables near the surface. Traditional statistical methods used at this scope \citep{Warner2010} have then been compared with NN which have displayed good results. For instance, \citet{rasp2018} compare several post-processing techniques on global ensemble forecast  to predict 2 m temperature at weather stations in Germany, using several forecasted fields. They found that the NN post-processing provides the best results for most stations. 
Other studies showed that using NN for NWP model post-processing is relevant for precipitation forecasting. For instance, \citet{frnda2022} and \citet{liu2023} improved global forecasts up to a lead time of 3~days, showing better results than with other post-processing methods and reducing the loss of predictability with lead time from the NWP.

As discussed above, NN which leverage observational data for rainfall predictions perform well for short lead times, but the prediction performance drops rapidly when considering longer forecast horizons. Rather than fully replacing NWP models with Deep Learning, an alternative approach, motivated by the success of NN for post-processing NWP, is to integrate NWP outputs in the NN input alongside with observations. 
Among the few studies that applied this solution, \citet{espeholt2022} used MetNet-2 for rainfall prediction for lead times up to 12~h. Using only observations or post-processing performs better at lead times shorter and longer than 6~h, respectively, while the hybrid approach, i.e. using both observed and NWP data, gives the best performance for all lead times.

This paper develops a hybrid approach for the prediction of rainfall threshold exceedance in the Mediterranean, in order to use the advantages of both observations and NWP data for stable rainfall predictions with lead times up to 24~h. 
The focus is on HPE, which both NWP and data-driven models struggle to forecast. In the case of data-driven models, the challenge for HPE is to learn to forecast events that are by definition rarely represented in the dataset from which the AI algorithm will learn. 
To counterbalance this, some studies suggested the use of a loss function, the function that compares AI predictions with reality during training, with weights to enhance training in favor of the least represented events \citep{shi2017, leinonen2022, liu2023}.
In this paper, a custom loss function based on the Peirce Skill Score (PSS), an equitable evaluation metric, is presented and compared with another more commonly used loss function and its weighted variant. 

This article is organized as follows. Section~\ref{sec:Data} presents the data used for this work, the performed forecasting tasks and their related NN models. The results are presented in section~\ref{sec:Results}, and the conclusions of this work is related in section~\ref{sec:Conclusions}. 
%
%
%
%
%
%
\section{Data and Methods}
\label{sec:Data}
%
%
%
%
%
%
\subsection{MeteoNet}
\begin{figure*}[h!]
    \centering
    \includegraphics[width=0.7\linewidth]{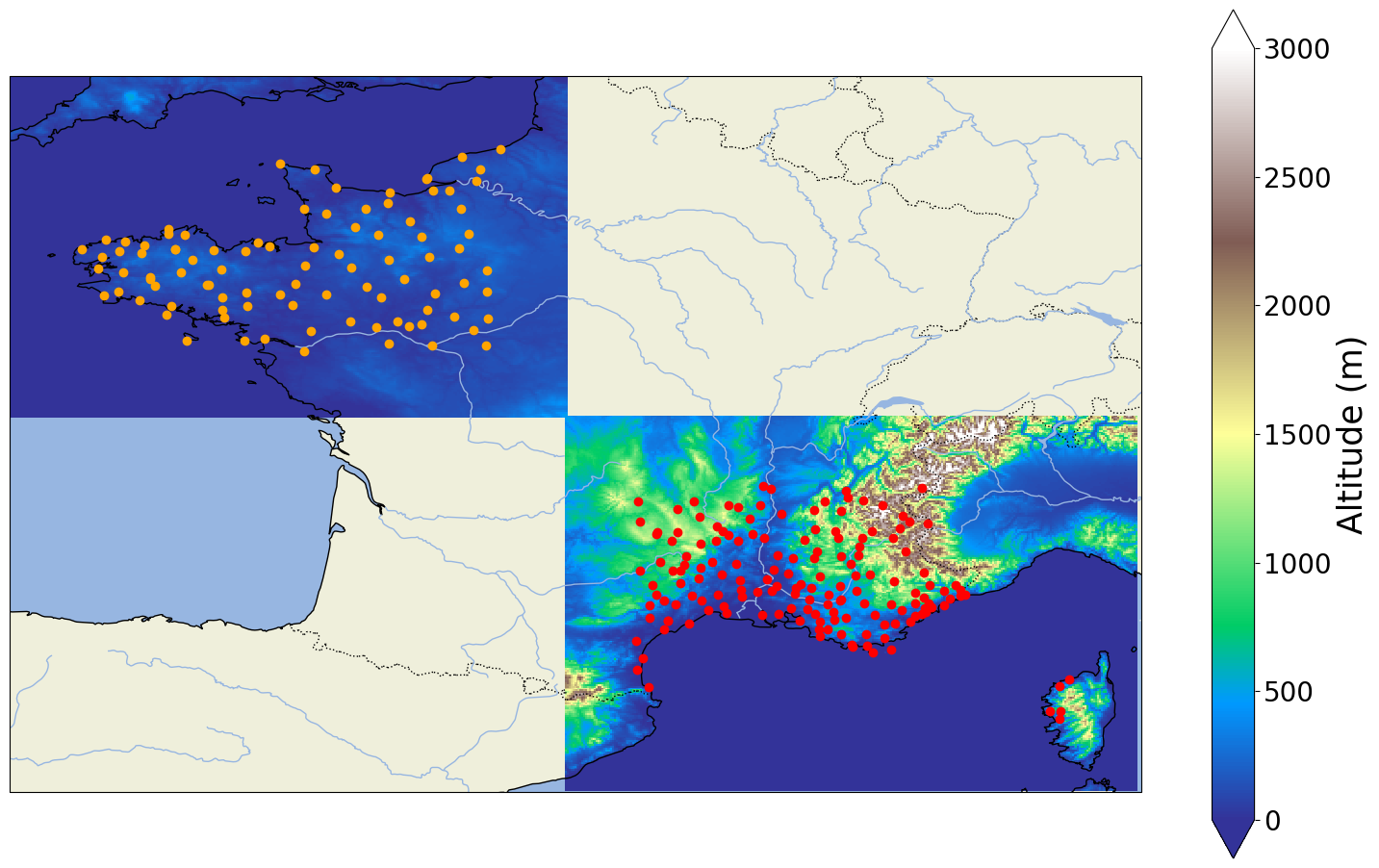}
    \caption{Geographical extent of the MeteoNet database, with ground station localization (in orange: North-West, in red: South-East) and topography in shade of color}
    \label{fig:sta_loc}
\end{figure*}

This study uses data from MeteoNet \citep{Meteonet}, an open source dataset provided by Météo France. This dataset includes many different type of weather data, both observations and NWP forecasts, that are available over two regions of France and over the 2016-2018 three year period. In this study, observation data from ground weather stations and NWP models forecast are used. Figure \ref{fig:sta_loc} shows the extent of NWP data and the localization of the ground stations. The two available regions are: North-West France, with 93 stations, and South-East France, with 159 stations. The number of ground-station has been reduced in this study, in comparison with the original dataset, due to the removal of missing values and meteorological fields. North-West and South-East regions are referred here respectively as ``oceanic'' region and ``Mediterranean'' according to Koppen's climate classification \citep{koppen}. Figure \ref{fig:sta_loc} highlights the difference between the two regions in terms of orography, with the Mediterranean region featuring high mountain ranges.

Observed meteorological fields at ground stations are provided with a time step of 6 min. The available NWP data include forecasts from two different Météo France operational models, Arpege \citep{Arpege} and Arome \citep{Arome}. Arpege is a global model, with a horizontal resolution of 0.1° (10 km). At this resolution, deep convection is not explicitly solved, and is therefore parametrized. Arome is a regional model that forecasts the weather in France with a higher spatial resolution of 0.025° (2.5 km), meaning that Arome is a convection-permitting model. The data available in MeteoNet comes from the operational model versions from 2016 to 2018, and are provided over the oceanic and Mediterranean regions portrayed in Figure \ref{fig:sta_loc}. For each day, and up to 24 hours ahead, only the forecast from the initialization at 00 UTC of both models are provided. Near surface fields from Arome are given with a time step of 1 h. Arpege forecasts are provided on 7 vertical levels, with the same extent as the surface fields, with a time step of 1 h between 00 UTC and 12 UTC and 3 h between 12 UTC and 24 UTC. To simplify the data pre-processing, this 3-dimensional Arpege data is selected with a constant time step of 3 h.
In this study, all meteorological fields available for this three type of data are used, that is, ground-station observations data, surface forecast fields from Arome and 3-dimensional forecast data from Arpege. The summary of the data used in this study is presented in Table \ref{tab:data}.

\begin{table*}[h!]
\centering
\caption{Summary of the data used in this study}
\begin{tabular}{ccc}
\hline\hline
Source  & \multicolumn{1}{c}{Spatial and temporal resolutions}                                                                                                                                                                                                                     & \multicolumn{1}{c}{Meteorological Fields}                                                                                                                                                                                                                                                                                       \\ \hline
Station observations & \begin{tabular}[c]{@{}c@{}}- Spatial resolution: ponctual \\- Time resolution: 6 min\end{tabular}                                                                                                                                                                   & \begin{tabular}[c]{@{}c@{}}- Wind direction (°)\\ - Wind speed (m~s$^{-1}$)\\ - Precipitation (mm)\\ - Humidity (\%)\\ - Dew point (K)\\ - Temperature (K)\\ - Mean sea level Pressure (Pa)\end{tabular}                                                                                                                          \\ \hline
NWP Arome   & \begin{tabular}[c]{@{}c@{}}- Spatial resolution: 0.025°\\ - Time resolution: 1 h\end{tabular}                                                                                                                                                                       & \begin{tabular}[c]{@{}c@{}}- 2 meters temperature (K)\\ - 2 meters dew point (K)\\ - 2 meters relative humidity (\%)\\ - 10 meters wind speed (m~s$^{-1}$)\\ - 10 meters wind direction (°)\\ - 10 meters u and v wind components (m~s$^{-1}$)\\ - Mean sea level pressure (Pa)\\ - Total precipitation since initialization (mm)\end{tabular} \\ \hline
NWP Arpege isobar levels & \begin{tabular}[c]{@{}c@{}}- Spatial resolution: 0.1°\\ with 7 isobar levels\\ (1000, 950, 925, 850, 700, 600,\\ 500 hPa)\\ - Time resolution: 3 h\end{tabular} & \begin{tabular}[c]{@{}c@{}}- Temperature (K)\\ - Wet bulb potential temperature (K)\\ - Relative humidity (\%)\\ - Wind speed (m~s$^{-1}$)\\ - Wind direction (°)\\ - u and v wind components (m~s$^{-1}$)\\ - Vertical velocity (Pa~s$^{-1}$)\\ - Geopotential (m$^2$~s$^{-2}$)\\ \end{tabular}                        \\ \hline
NWP Arpege height levels & \begin{tabular}[c]{@{}c@{}}- Spatial resolution: 0.1°\\ with 7 vertical levels\\ (20, 100, 500, 875, 1375, 2000,\\ 3000 m)\\ - Time resolution: 3 h\end{tabular} & \begin{tabular}[c]{@{}c@{}}- Pressure (Pa)\end{tabular}                        \\ \hline
\end{tabular}

\label{tab:data}
\end{table*}

%
%
%
%
%
\subsection{Forecasting task}

The forecasting task in this study is defined as follows: Given a point in time, $t$, referred to as the NN-model initialization time, and a specific ground station, the objective is to determine whether the accumulated precipitation between $t$ and $t+\Delta t$ (hereafter referred to as $R_{\Delta t}(t)$) will be lower or greater than a given rainfall threshold $Q_{\Delta t}$. The studied values of $\Delta t$, defined as the accumulation window size (hereafter referred as window size), are of 1~h, 2~h, 3~h, 6~h, 12~h, and 24~h.
As this task amounts to forecast two classes of rain---below or above $Q_{\Delta t}$---, lets define $Y_{t+\Delta t}$ as the corresponding rainfall class such as : 

\begin{equation}
\label{eq:for_task}
\left\{
    \begin{array}{ll}
        Y_{t+\Delta t}=0~\text{ if }~R_{\Delta t}(t) < Q_{\Delta t}
        \\
        Y_{t+\Delta t}=1~\text{ if }~R_{\Delta t}(t) \geq Q_{\Delta t} %
    \end{array}
\right.
\end{equation}

A machine learning approach for classification task predicts the probability $p_{t + \Delta t}$ that $Y_{t+\Delta t}=1$. A critical probability $P_C$ must then be considered, such as, with $\widehat{Y}_{t+\Delta t}$ the predicted rain class : 

\begin{equation}
\label{eq:for_task_2}
\left\{
    \begin{array}{ll}
    \widehat{Y}_{t+\Delta t} = 0~\text{ if }~p_{t + \Delta t} < P_C
    \\
    \widehat{Y}_{t+\Delta t} = 1~\text{ if }~p_{t + \Delta t} \geq P_C
    \end{array}
\right.
\end{equation}

As explained in \cite{Jolliffe2004} and \cite{WILKS2019369}, an optimal value \( P_C^\star \) can be selected for \( P_C \) to optimize the chosen score for assessing prediction quality (more details in subsection~\ref{sec:Data}–\ref{subsec:metrics}).

The prediction uses $\mathcal{X}$ which represents the NN-model inputs. All data showed at Table \ref{tab:data} are used as inputs: station inputs are defined as the observations at the given station from $t - \Delta t$ to $t$, with a time step of 6~min, and Arome and Arpege inputs are defined as the prediction of each NWP model with 10x10 grid points around the given station from $t$ to $t + \Delta t$, the former with a time step of 1~h, the latter with a time step of 3~h. The learning target---i.e. the data from which the NN-model learns from---are calculated using the accumulated rainfall between $t$ and $t + \Delta t$ observed at the ground station. More details are given in subsection~\ref{sec:Data}-\ref{NNhybdt} regarding data management for training, and the choice of the value of $Q_{\Delta t}$ is discussed in subsection~\ref{sec:Data}-\ref{subsec:Q}.

\begin{figure*}[h!]
    \centering
    \includegraphics[width=\linewidth]{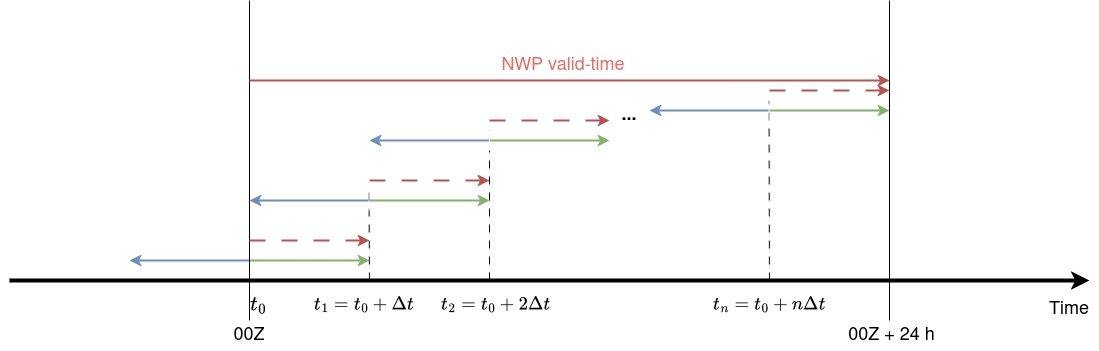}
    \caption{Diagram of forecasting tasks during a day depending on the value of $\Delta t$. At each $t_j$, a prediction is made by the NN-model. Blue arrows represent the time period of the observed data used for the NN-model prediction. Dashed red arrows represent the time period of the NWP models predictions data used for the NN-model prediction. The red arrow represent the time period of NWP models predictions from their initialization. Green arrows represent the time period of $\Delta t$}
    \label{fig:flex_exp}
\end{figure*}

Figure \ref{fig:flex_exp} summarizes the forecasting task for different NN-model initialization times $t_j$ during a given day. Index $j=1,\dots,n$ runs over the number of tasks per day, which is $n=\frac{24}{\Delta t}$ for $\Delta t \geq 3~h$ and $n=8$ for $\Delta t < 3~h$ to match Arpege time step of 3~h and avoid overlapping experiments. 
For the NN-model initialization time $t_j$, $R_{\Delta t}(t_j)$ (green arrows) is predicted using either station data observed between $t_j - \Delta t$ and $t_j$ (blue arrows), NWP models data predicted between $t_j$ and $t_j + \Delta t$ (dashed red arrows), or both. Depending on the accumulation window size $\Delta t$ value (hereafter referred as window size), given that MeteoNet provides NWP forecast data only from the 00 UTC initialization run (top red arrow), the NN-model may use NWP data with a greater or lesser temporal distance from NWP model initialization. This fact should be kept in mind when discussing the obtained results.

It is important to note that the number of experiments varies depending on the value of $\Delta t$, as $\Delta t$ also represents, for $\Delta t \geq 3~h$, the temporal distance between two NN-model initialization. One can estimate the number of experiments $N_{exp}$ such as $N_{exp} = n \times 365 \times 3 \times N_{station}$, with $N_{station}$ the number of weather stations considered, $3 \times 365$ representing the number of days during the three studied years. Hence, as the proportion of experiments being discarded due to missing values is around 10\% for each experiments, in the Mediterranean region, the number of experiments with $\Delta t = 24~h$ and $\Delta t = 3~h$ is respectively $N_{exp} \approx 150~000$ and $N_{exp} \approx 1~200~000$.
Hereafter index $i=1, \dots,N_{exp}$ will be used to address the experiments and $p_i$ will denote the value $p_{t+\Delta t}$ of the model at indice i.

%
%
%
%
%
%
\subsection{Rainfall threshold values}
\label{subsec:Q}

For each region and each $\Delta t$, if $R_{\Delta t}(t)$ represents the cumulative precipitation observed at some weather station between $t$ and $t+\Delta t$, the ''climatological distribution'' allows one to define the probability that $R_{\Delta t}$ is above or below a given threshold $z$. For instance, the so-called Complementary Cumulative Distribution Function (CCDF) of the climatology can be defined as the probability of exceedance of a given rainfall value:
$$  
       P_{\Delta t}(z) = \text{Probability} \Big[ R_{\Delta t} \geq z \Big] = \frac{\text{Numb. of events} \;  \{R_{\Delta t} (t) \geq z\}}{N_{exp}} 
$$

\begin{figure*}[h!]
    \centering
    \begin{subfigure}{0.33\linewidth}
        \centering
        \caption{}
        \includegraphics[width=0.9\textwidth]{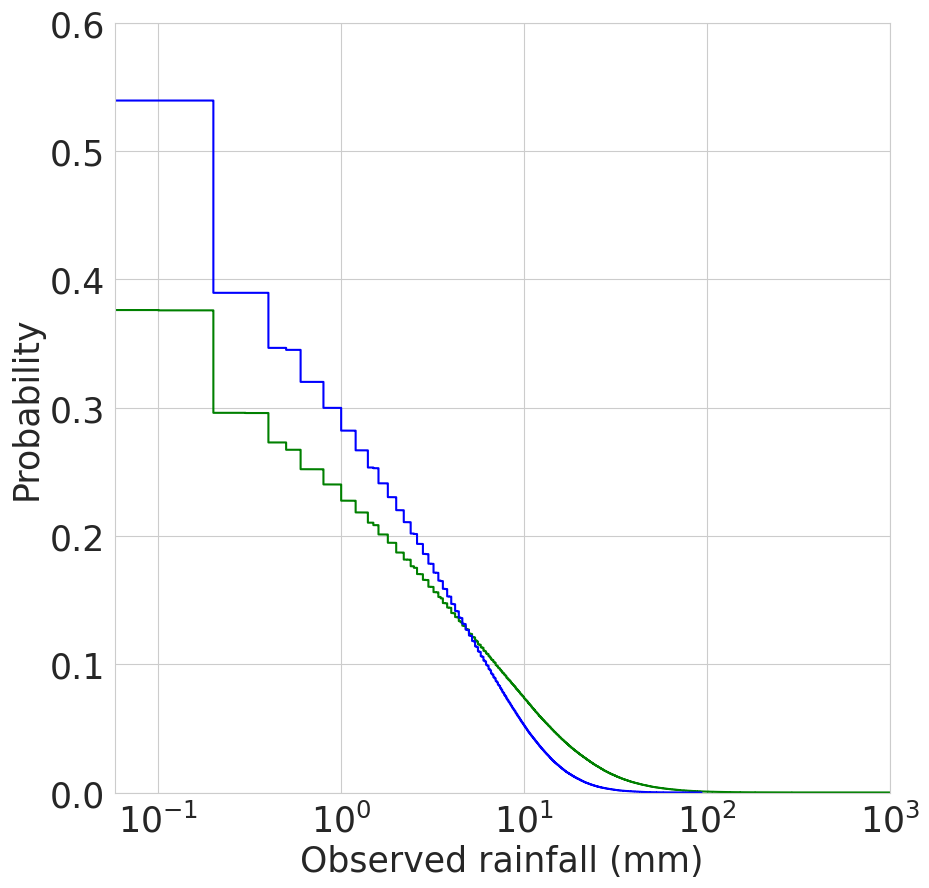}
        \label{fig:ccdf}
    \end{subfigure}%
    \begin{subfigure}{0.33\linewidth}
        \centering
        \caption{}
        \includegraphics[width=0.9\textwidth]{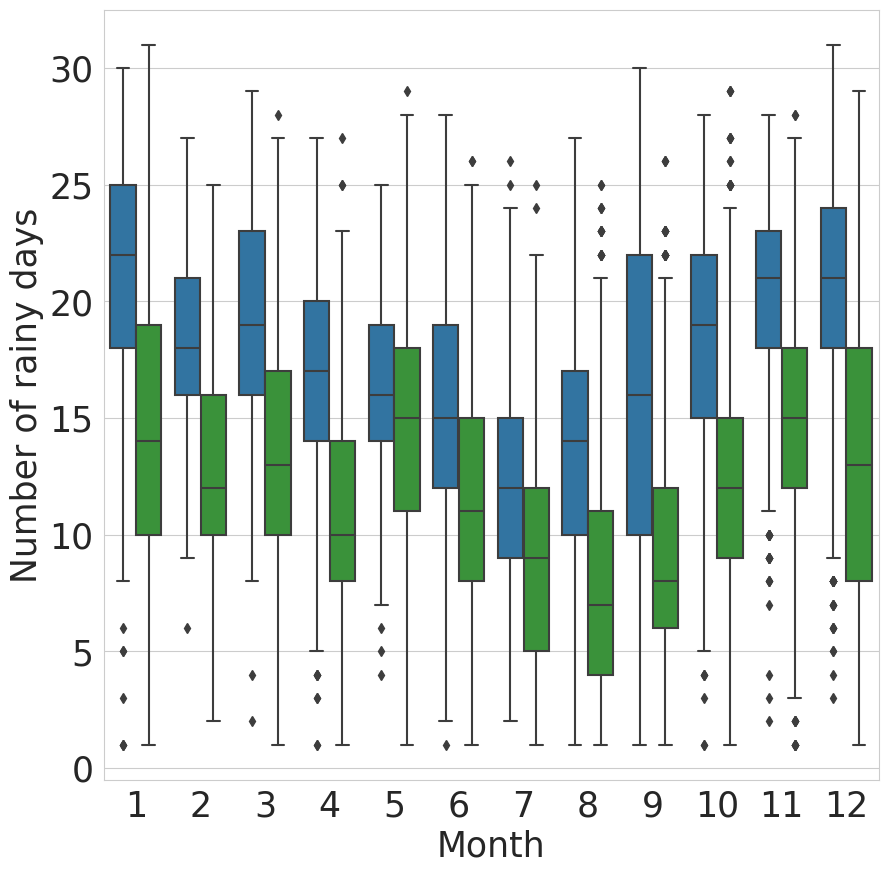}
        \label{fig:rainy_day}
    \end{subfigure}
    \begin{subfigure}{0.33\linewidth}
        \centering
        \caption{}
        \includegraphics[width=0.9\textwidth]{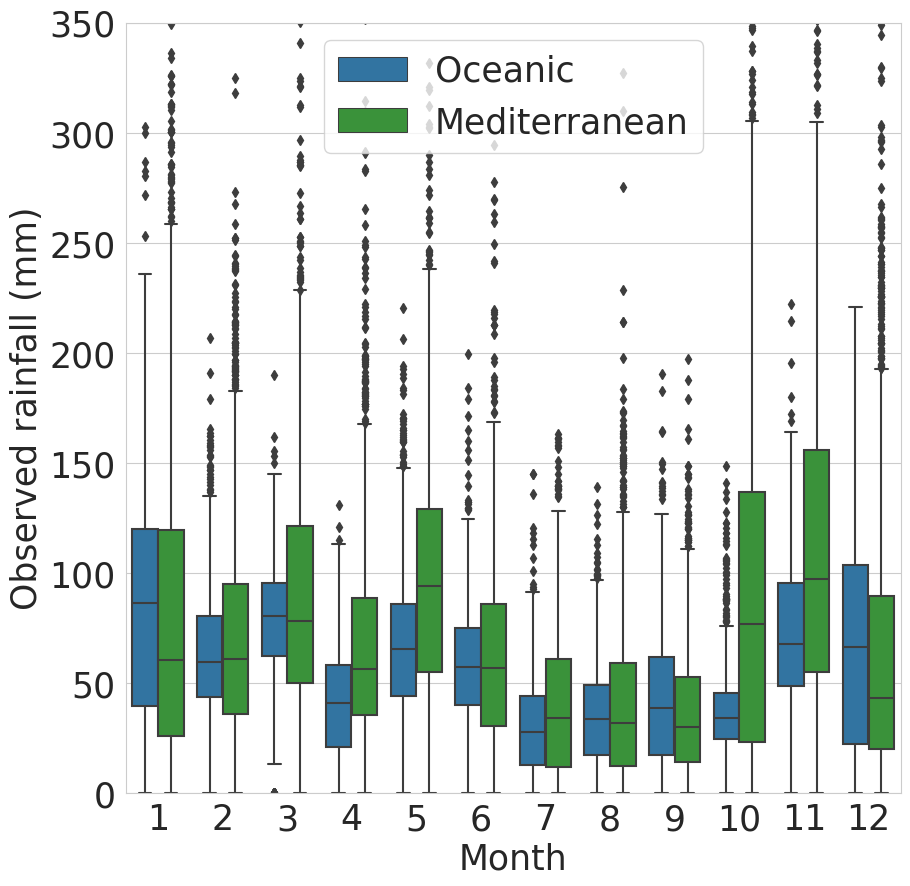}
        \label{fig:total_year}
    \end{subfigure}
    \caption{(a) Complementary Cumulative Distribution Function of the climatology for $\Delta t = 24~h$. (b) Boxplot of number of rainy days by month and region. (c) Boxplot of monthly cumulated rainfall by month and region. In blue: oceanic region, in green: Mediterranean region. }
    \label{fig:climato}
\end{figure*}

Figure \ref{fig:climato} summarizes the differences between the two regions by illustrating the climatology CCDF for $\Delta t = 24~h$. This Figure highlights the fact that low to moderate rain is more likely to occur in the oceanic region, while heavier rain are more frequent in the Mediterranean. Note that the value at $z<10^{-1}$ directly indicates the probability at observing a rainy day. This result can also be seen in Figure \ref{fig:rainy_day} that shows the number of rainy days (number of days for which the recorded daily precipitation accumulation was above 0 mm) for each month by regions: one is more likely to observe rain in the oceanic region than in the Mediterranean. However, even though more rainy days are found in the oceanic region, Figure \ref{fig:total_year} reveals that the monthly cumulated rainfall is generally higher in the Mediterranean region. This is particularly true during fall (especially October and November) which is when most HPE occur in the Mediterranean \citep{Ricard2012}.

Reference thresholds $Q_{\Delta t}$ for each region can be simply defined from $P_{\Delta t}(z)$ as the values of $z$ associated with a given set of probability level $s$. These are the so-called ``quantiles''. 
By definition, $QL$, the quantile of level $L$, is the value $z$ such
that:
\begin{equation}
\label{eq:def_Q}
 P_{\Delta t}(z=QL) = s \; \text{with} \; s = 1-\frac{L}{100} \; . 
\end{equation} 
For example the $95^{th}$ quantile, $Q95$, is associated with $s= 0.05$ and is such that $P_{\Delta t}(Q95) = 0.05$.
Note that the smaller the value of \( s \) (i.e., the larger \( L \)), the rarer and more severe the considered events become.  

Table \ref{tab:quantiles} summarizes the values of the 95\textsuperscript{th}, 99\textsuperscript{th}, 99.5\textsuperscript{th}, and 99.9\textsuperscript{th} quantiles of the climatology for each region and $\Delta t$. These quantiles, which correspond to increasingly intense events, serve as the thresholds $Q_{\Delta t}$ used throughout this study to classify cumulative precipitation events into class 0 or 1 (i.e., $Y = 0$ or $Y = 1$ in Eq. \eqref{eq:for_task}). By definition, the relative frequency of the class 1 event ($Y=1$) is $s$, namely it will be $s=0.05$ when considering $Q_{\Delta t}=Q95$, $s=0.01$ for $Q_{\Delta t}=Q99$ and so on.

\begin{table*}[h!]
\centering
\caption{Values of quantiles for each region and each $\Delta t$. Values are in $mm$.}
\begin{tabular}{ccccccccc}
\hline \hline
                                    & \multicolumn{2}{c}{\textbf{Q95}}              & \multicolumn{2}{c}{\textbf{Q99}}              & \multicolumn{2}{c}{\textbf{Q99.5}}            & \multicolumn{2}{c}{\textbf{Q99.9}}            \\ 
                                   $\Delta t$ & \multicolumn{1}{c}{\textit{Oceanic}} & \textit{Mediterranean} & \multicolumn{1}{c}{\textit{Oceanic}} & \textit{Mediterranean} & \multicolumn{1}{c}{\textit{Oceanic}} & \textit{Mediterranean} & \multicolumn{1}{c}{\textit{Oceanic}} & \textit{Mediterranean} \\ \hline
\multicolumn{1}{c}{\textbf{1 h}}  & \multicolumn{1}{c}{0.4}         & 0.4         & \multicolumn{1}{c}{1.8}         & 2.6         & \multicolumn{1}{c}{2.8}         & 4.0         & \multicolumn{1}{c}{5.5}         & 9.1         \\ 
\multicolumn{1}{c}{\textbf{2 h}}  & \multicolumn{1}{c}{0.8}         & 0.8         & \multicolumn{1}{c}{3.4}         & 5.0         & \multicolumn{1}{c}{4.9}         & 7.6         & \multicolumn{1}{c}{9.2}         & 15.7        \\ 
\multicolumn{1}{c}{\textbf{3 h}}  & \multicolumn{1}{c}{1.4}         & 1.4         & \multicolumn{1}{c}{4.8}         & 7.2         & \multicolumn{1}{c}{6.8}         & 10.8        & \multicolumn{1}{c}{12.3}        & 21.3        \\ 
\multicolumn{1}{c}{\textbf{6 h}}  & \multicolumn{1}{c}{2.8}         & 3.3         & \multicolumn{1}{c}{8.4}         & 13.1        & \multicolumn{1}{c}{11.2}        & 18.6        & \multicolumn{1}{c}{18.7}        & 34.7        \\ 
\multicolumn{1}{c}{\textbf{12 h}} & \multicolumn{1}{c}{5.6}         & 7.2         & \multicolumn{1}{c}{13.5}        & 22.3        & \multicolumn{1}{c}{17.4}        & 30.4        & \multicolumn{1}{c}{27.8}        & 55.4        \\ 
\multicolumn{1}{c}{\textbf{24 h}} & \multicolumn{1}{c}{10.3}        & 14.1        & \multicolumn{1}{c}{20.2}        & 35.8        & \multicolumn{1}{c}{24.8}        & 48.5        & \multicolumn{1}{c}{38.2}        & 89.6        \\
\hline
\end{tabular}
\label{tab:quantiles}
\end{table*}

Therefore, the classification task is performed for a highly unbalanced dataset and the approach to this issue is discussed in subsections \ref{sec:Data}-\ref{subsec:metrics} and \ref{sec:Data}-\ref{subsec:loss}. 
%
%
%
%
%
%
\subsection{Neural-Network to handle hybrid data}\label{NNhybdt}

\begin{figure*}[h!]
    \centering
    \includegraphics[width=\linewidth]{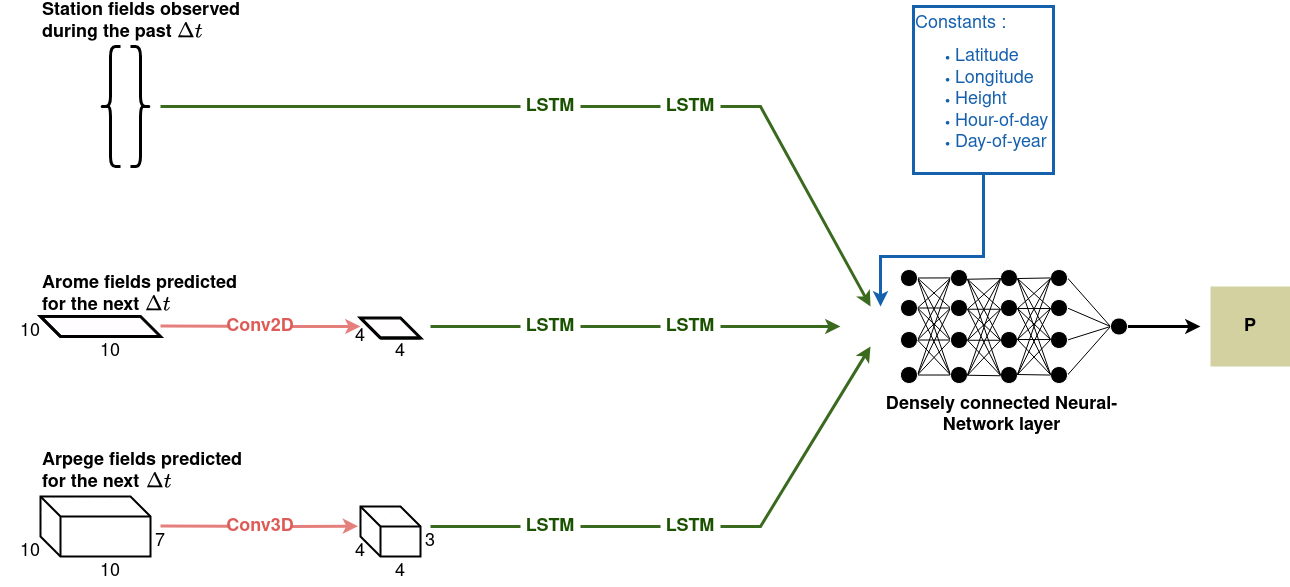}
    \caption{Neural-Network architecture. The top branch represents how data from stations are processed. The middle branch represents how Arome predicted surface fields data are processed. The bottom branch shows how the 3-dimensional field data predicted by Arpege are processed. Green arrows represent temporal processing of data, red arrows represent spatial processing of data. If the different types of data are used together, the output of each branch is concatenated. Then, the output is passed through a densely connected NN. The output of this NN architecture is the probability $p$  of occurrence of class 1. }
    \label{fig:neural-network}
\end{figure*}

Figure \ref{fig:neural-network} illustrates the NN architecture used in this study and highlights how the different data sources are integrated by the NN-model during training. This architecture answers one of the following prediction tasks, with $\mathcal{M}$ the NN-model output:

\begin{equation}
    \centering
    \left\{
    \begin{array}{ll}
        p_{t+\Delta t} = \mathcal{M}(\mathcal{X}^{Station}_{t,t-\Delta t}) \\
        p_{t+\Delta t} = \mathcal{M}(\mathcal{X}^{Arome}_{t,t+\Delta t}) \\
        p_{t+\Delta t} = \mathcal{M}(\mathcal{X}^{Arpege}_{t,t+\Delta t})\\
        p_{t+\Delta t} = \mathcal{M}(\mathcal{X}^{Hybrid}_{t-\Delta t,t+\Delta t})
    \end{array}
\right. 
\end{equation}

With $\mathcal{X}^{Station}_{t,t-\Delta t}$ the 7 recorded meteorological fields at the station during the past $\Delta t$ hours, $\mathcal{X}^{Arome}_{t,t+\Delta t}$ the 8 forecasted weather fields by Arome on 10x10 grid points around the ground-station during the next $\Delta t$ hours, $\mathcal{X}^{Arpege}_{t,t+\Delta t}$ the 9 forecasted fields by Arpege on 10x10 grid points around the ground-station with 7 vertical levels during the next $\Delta t$ hours, and $\mathcal{X}^{Hybrid}_{t-\Delta t,t+\Delta t} = \{ \mathcal{X}^{Station}_{t,t-\Delta t}, \mathcal{X}^{Arome}_{t,t+\Delta t}, \mathcal{X}^{Arpege}_{t,t+\Delta t} \}$, defined as the hybrid dataset. 

Given the time-steps of respectively 6 min, 1 h and 3 h, $\mathcal{X}^{Station}_{t,t-\Delta t}$ shape is ($6 \times \Delta t,7$), $\mathcal{X}^{Arome}_{t,t+\Delta t}$ shape is ($\Delta t,10,10,8$) (Arome forecasts at time $t$ being excluded during data pre-processing), and $\mathcal{X}^{Arpege}_{t,t+\Delta t}$ shape is ($\frac{\Delta t}{3} +1,10,10,7,9$) for $\Delta t \geq 3~h$ and ($1,10,10,7,9$) for $\Delta t < 3~h$.

Long-Short Term Memory (LSTM) neural-networks are used to extract the temporal information from the feature data. 
Arome and Arpege data, which consist of 2D and 3D representations of atmospheric variables, respectively, are first processed by a Convolutional neural-network (Conv2D and Conv3D) to extract spatial information before being passed to an LSTM. 
In alignment with the assigned forecasting task, the output layer is a fully-connected NN with a Sigmoid activation function, whose output $p_{t+\Delta t}$ is the probability of class 1 occurrence.
A series of experiments is conducted in which each type of input data (Station, Arome and Arpege) is considered and analyzed separately. Additionally, the combination of each input data in a hybrid dataset is considered, in which case the respective NN structures of each dataset are concatenated together. Finally, ground-station spatial and temporal information data (latitude, longitude and height of the ground-station as well as hour of the day and day of the year) are concatenated with each structure before passing through a Fully-Connected NN. Hereafter, NN-model trained with either Station, Arome, Arpege or Hybrid dataset will be called respectively NN-Station, NN-Arome, NN-Arpege and NN-Hybrid.

 Data are randomly distributed between the training, validation and test dataset, with a respective ratio of $80\%$, $15\%$ and $5\%$, taking care to use separate days when building the three datasets to avoid biases in the predictions.

%
%
%
%
%
%

\subsection{Evaluation scores}
\label{subsec:metrics}
In this study, forecasts are treated as deterministic. This means that model evaluation is based directly on the binary outcomes $\widehat{Y}_{i}$ corresponding to model outputs $p_i$ by means of Eq. \eqref{eq:for_task_2}, rather than on the $p_i$ themselves. Despite the relative simplicity of this approach, evaluating a binary classification task is not straightforward, as various evaluation scores can be used. These scores are derived from the components of the confusion matrix, that is, a 2×2 Table comparing predicted outcomes $\widehat{Y}_i$ and actual outcomes $Y_i$ using four key components: true positives ($TP$), false positives ($FP$), false negatives ($FN$) and true negatives ($TN$).  
Naturally, $TP+FP+FN+TN=N_{exp}$ where $N_{exp}$ is the total number of events and $\frac{TP+FN}{N_{exp}} = s$ where is the level of probability associated with the chosen quantile threshold $Q_{\Delta t}$ (see \eqref{eq:def_Q}).  Considering this, the problem of evaluating binary forecasts-observation pairs $\left( \widehat{Y}_i,Y_i\right)$  is inherently a three-dimensional problem. When using a scalar score for forecast evaluation, some information is inevitably lost. Therefore, the choice of this scores should be guided by the specific characteristics of the problem being analyzed. 
Since the classification task considered here concerns the prediction of rare events, this study argues that forecast performances should be evaluated under an equitable skill score. Indeed, following \citet{gandin1992equitable}, equitable metrics not only give equal value to random and constant forecasts, but also weight more strongly correct forecasts of rare events, which discourages artificial distortions towards the more common event. 
Taking this into consideration, this study uses the Peirce Skill Score \citep{PSS} (also called True Skill Statistic or Hanssen-Kuipers Discriminant) to evaluate HPE prediction. 
This score is the only equitable metric for binary classification which strictly follows the equitability criteria introduced in \citet{gandin1992equitable} and is independent from the class relative frequency $s$ and thus particularly suitable to evaluate predictions of rare events \citep{Woodcock1976,Rodwell2011,Ebert2022}.
PSS is positively oriented and its possible values ranges from -1 to 1, 0 meaning no skill and 1 perfect skill. It is defined as the difference between the \emph{Hit Rate} $H=\frac{TP}{TP+FN}$ and the \emph{False Alarm Rate} $F= \frac{FP}{TN+FP}$: 

\begin{equation}
\label{eqPSS}
    \centering
    PSS = H - F = \frac{TP}{TP+FN} - \frac{FP}{TN+FP}\,.   
\end{equation}

The Hit-Rate H, also known as Recall, is often used together with Precision, defined as $Precision = \frac{\text{TP}}{\text{TP} + \text{FP}}$  to evaluate classification performance on unbalanced datasets.
Both Precision and Recall values ranges from 0 to 1, 0 meaning no skills and 1 perfect skill. Precision measures the likelihood that a predicted class 1 event is actually observed, while Recall quantifies the probability that an observed class 1 event was correctly forecasted. The comparison between these two metrics illustrates the trade-off between accurately identifying events and minimizing missed detections.

For completeness, the results obtained for two additional metrics are considered, the Critical Success Index (CSI), which is also frequently used in case of class unbalance, and the Heidke Skill Score \citep{Jolliffe2004,WILKS2019369}. Definition of such scores, as well as the relative model results, are reported in Appendix~B.   


\subsection{Loss function}
\label{subsec:loss}
The choice of the loss function is a crucial aspect of the learning process. The loss function guides the NN through the comparison with the ground-truth, and the focus of the NN training will change depending on the loss function. 
In this study, a comparison of three different loss functions is performed. The first one, the \emph{Binary Cross-Entropy} (BCE), is the loss function classically used for the classification of binary events and it is defined as: 
        \begin{equation}
    \label{eq:bce}
        \centering
        BCE = -\frac{1}{N}\sum_{i=1}^N Y_i \log(p_i) + (1 - Y_i) \log(1 - {p_i})\,,
    \end{equation}
where $Y_i$ is the ground-truth and ${p_i}$ the predicted probability at prediction task indice i.

As this study focuses on rare events, two loss functions that explicitly account for the unbalanced dataset are also experimented.
These are the \emph{Weighted Binary Cross-Entropy} (w-BCE) and the \emph{Peirce Loss} (PL).
w-BCE is a variant of BCE in which weights $w_1$ and $w_0$ are introduced to handle the unbalance of the relative classes: 
    
    \begin{equation}
    \label{eq:wbce}
        \centering
        \text{w-}BCE = -\frac{1}{N}\sum_{i=1}^N w_1 Y_i \log({p_i}) + w_0 (1 - Y_i) \log(1 - {p_i})\,.
    \end{equation}
The value of the $w_i$ is given by the frequency of the corresponding classes, that is $w_1=\frac{1}{s}$, $w_0=\frac{1}{1-s}$, $s$ being the mean observed rate of the HPE event.      
    
The Peirce Loss PL is a custom loss function 
that is straightforwardly derived from the PSS score (\ref{eqPSS}): 
    
    \begin{equation}
   \label{eq:pssloss}
        \centering
        PL = \sum_{i=1}^N {p_i} \frac{1 - Y_i}{\sum_{j=1}^N (1 - Y_j)}  - {p_i}  \frac{Y_i}{\sum_{j=1}^N Y_j}  
    \end{equation}
Indeed, by expressing the components of the confusion matrix in terms of forecasts $\widehat{Y}_i$ and observations $Y_i$, H and F are given by:
\begin{equation}
    H=\frac{\sum_i^N \widehat{Y}_i {Y}_i}{\sum_j^N {Y}_j}\qquad F=\frac{\sum_i^N \widehat{Y}_i (1-{Y}_i)}{\sum_j^N (1-{Y}_j)}\,,
       \label{eqHS}
\end{equation}
and then (\ref{eq:pssloss}) is obtained by setting $\widehat{Y}_i\approx p_i$.
It should be noted that when using a balanced loss function, more weight is assigned to the underrepresented class 1, leading to an expected increase in false positives (FP) and a decrease in false negatives (FN). This effect is not considered to be problematic, as the societal cost of over-forecasting HPEs is believed significantly lower than that of under-forecasting such events.

A discussion can be found in Appendix~A about the problem of the relationship between the ``true probability'' $q_{i}$ that $Y_{i} = 1$ occurs and $p_{i}$, the output of NN-model, according to the optimized loss. 

%
%
%
%
%
%
\section{Results}
\label{sec:Results}
%
%
%
%
%
%
\subsection{The benefits of the hybrid dataset}\label{sub_BHD}


The relative performances, in terms of PSS, obtained using  NN-Station, NN-Arome, NN-Arpege as well as the NN-Hybrid model are illustrated on Figure \ref{fig:PSS&HSS}. Results using alternative scores are presented in Appendix B.
The loss function chosen here is the Peirce Loss (PL), and further comparison between loss functions are discussed in section 3.b.
Performances obtained using a benchmark forecast, given by raw Arome predictions, are shown as well.
\begin{figure*}[h!]
\centering
\begin{subfigure}{.4\textwidth}
  \centering
    \caption{}
  \includegraphics[width=0.9\linewidth]{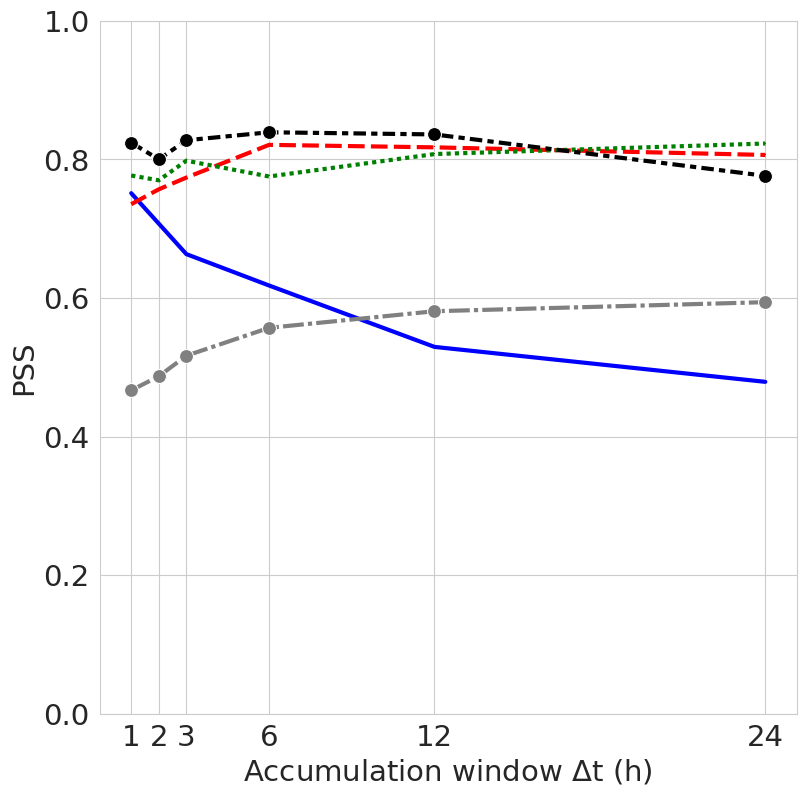}
  \label{fig:PSSvsdeltat}
\end{subfigure}
\begin{subfigure}{.4\textwidth}
  \centering
  \caption{}
  \includegraphics[width=0.9\linewidth]{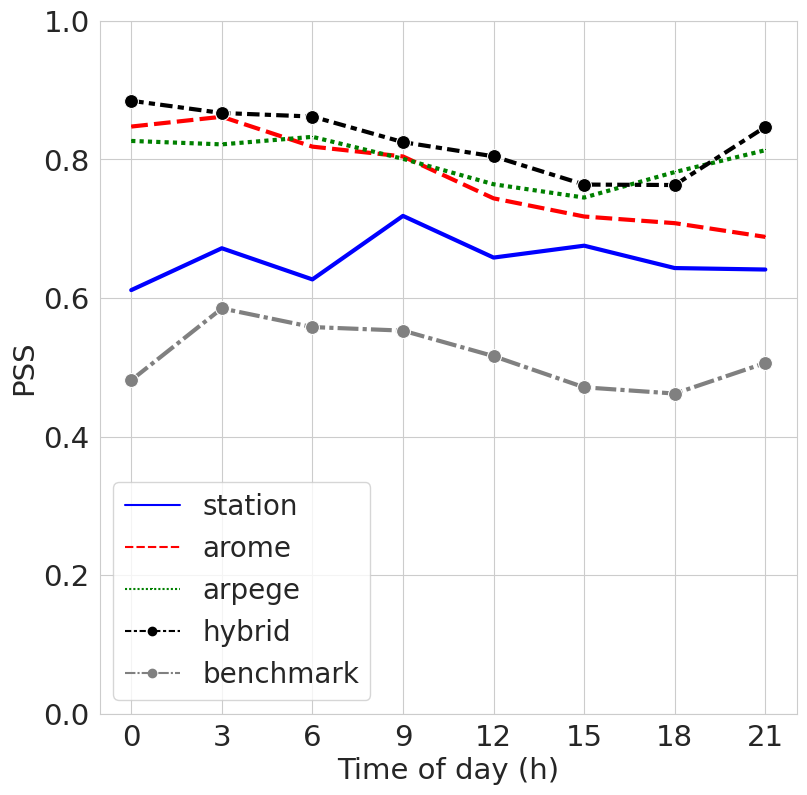}
  \label{fig:PSSvsvalidtime}
\end{subfigure}
\caption{PSS performance of NN-model trained on different types of input data from the Mediterranean region, for the 5\% rarest rainfall events ($95^{th}$ quantile) prediction, with the Peirce loss as loss function and $P_C = 0.5$. (a) Sensitivity to the window size $\Delta t$. (b) Sensitivity to the time of the day for a 3 hour window size. In solid blue: NN-Station. In dashed red: NN-Arome. In dashed green: NN-Arpege. In dotted black: NN-Hybrid. In dotted and dashed grey: benchmark.}
\label{fig:PSS&HSS}
\end{figure*}


Figure \ref{fig:PSSvsdeltat} displays the PSS for the 95\% quantile threshold exceedance relative to the window size $\Delta t$. 
The benchmark (dashed gray line) shows increasing PSS with $\Delta t$, with PSS value between $0.4$ and $0.5$ for $\Delta t = 1~h$ to PSS value around $0.6$ for $\Delta t = 24~h$. This means that the benchmark performs better at predicting rainfall threshold exceedance over a longer window size than over a shorter one.  
For NN-Arome (dashed red line), the behavior is similar to that of the benchmark in the sense that performances are better for longer $\Delta t$, but with much better performance overall: e.g., for $\Delta t = 24~h$, $PSS = 0.8$ instead of $PSS = 0.6$ for the benchmark.
NN-Arpege (dotted green line) shows PSS comparable with NN-Arome, although performs better for $\Delta t \leq 3~h$, thus providing a more stable performance with $\Delta t$ overall.
In contrast, the NN-Station model (solid blue line) better predicts rainfall threshold exceedance for short $\Delta t$ than longer one. It shows a rapid decrease in PSS with $\Delta t$, with PSS values higher than $0.6$ for $\Delta t \leq 6~h$---then outperforming the benchmark over this range of $\Delta t$ values---and decreasing below $0.6$ thereafter.
Finally the NN-Hybrid model (dotted black line) offers the best overall performance, with the highest PSS values and the most stability with $\Delta t$. Indeed, PSS values for short $\Delta t$ are similar to values for longer $\Delta t$, though a slight decrease can be seen for $\Delta t = 24~h$. 
This emphasizes the complementarity of observed and predicted data.

Figure \ref{fig:PSSvsvalidtime} shows the sensitivity of NN-models and benchmark predictions to the time of day at which the prediction is performed for $\Delta t = 3~h$. 
The benchmark depicts a PSS slightly sensitive to the time of day, with a PSS increase between 00 UTC and 03 UTC, followed by a slow decrease between 03 UTC and 18 UTC.
Though a similar trend can be seen for NN-Arome, PSS values are much higher overall, as seen in Figure \ref{fig:PSSvsdeltat}.
NN-Arpege provides a slightly decreasing PSS with time of day, although slower than NN-Arome and with a slight increase for time of day greater than 15~h. NN-Arpege outperforms NN-Arome after 6~h.
On the other hand, apart for few variations, NN-Station shows low sensitivity to the time of day. 
The differences in PSS sensitivity with time of day between NN-Station, NN-Arome and NN-Arpege can be explained by the fact that NN-Station use observed information that are continuously updated, whereas NN-Arome and NN-Arpege use data that are derived solely from NWP initialization at 00 UTC.
NN-Hybrid again delivers the best PSS performance. A slight sensitivity with time of day is portrayed, with slowly decreasing PSS followed by an increase for 21 UTC. This dataset improves the NN-model's performance on the PSS overall and mitigates time of day sensitivity.

%
%
%
%
\subsection{About the choice of the loss function}
\begin{figure*}[h!]
\centering
\begin{subfigure}{.39\textwidth}
  \centering
    \caption{}
  \includegraphics[width=\linewidth]{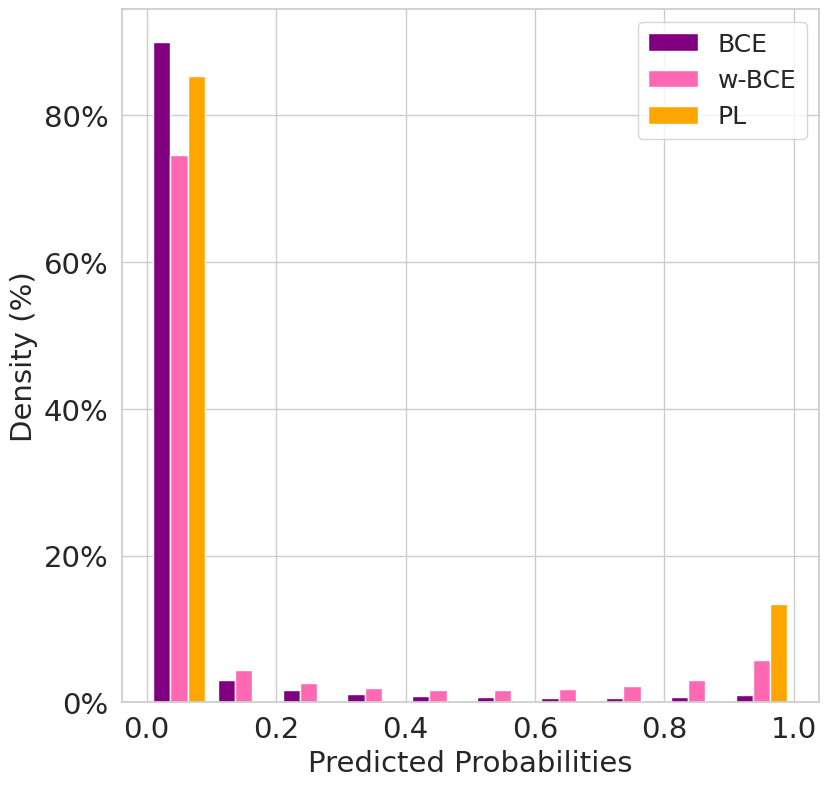}
  \label{fig:LossandPCimpact_a}
\end{subfigure}
\begin{subfigure}{.39\textwidth}
  \centering
  \caption{}
  \includegraphics[width=\linewidth]{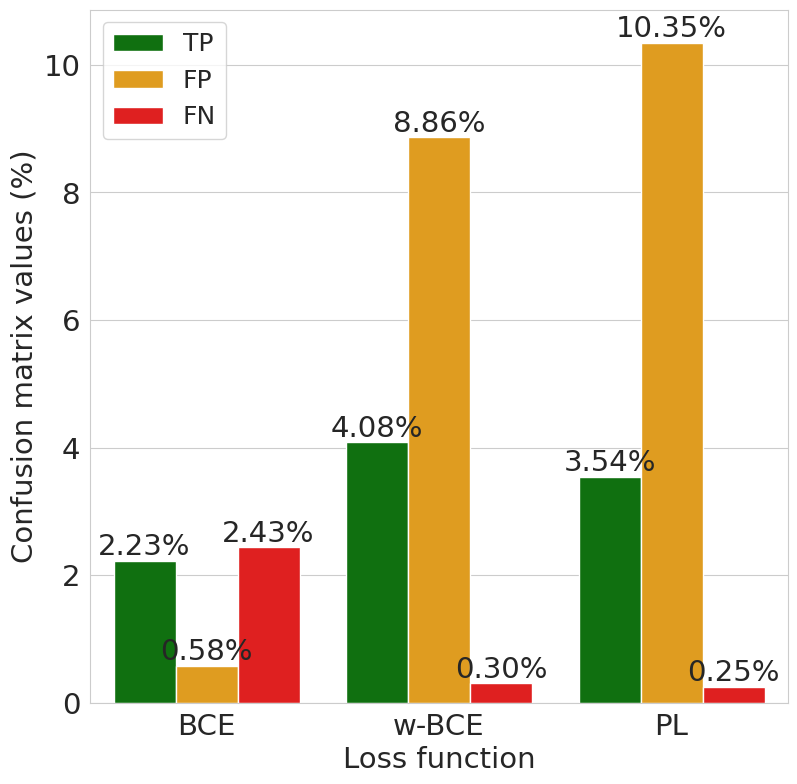}
  \label{fig:LossandPCimpact_b}
\end{subfigure}

\begin{subfigure}{0.39\linewidth}
    \centering
    \caption{}
    \includegraphics[width=\linewidth]{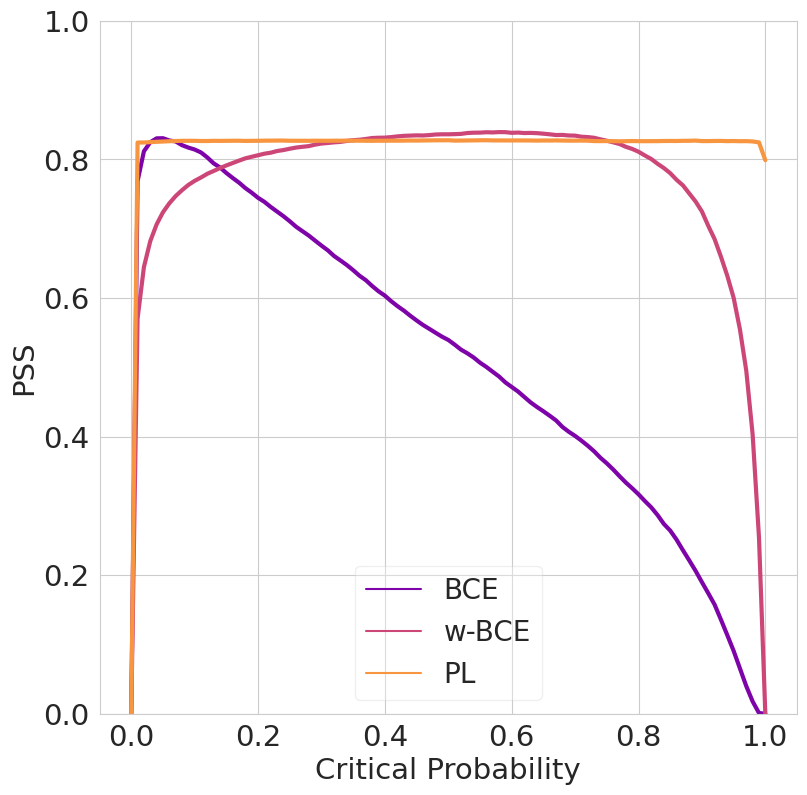}
    \label{fig:LossandPCimpact_c}
\end{subfigure}
\begin{subfigure}{0.39\linewidth}
    \centering
    \caption{}
    \includegraphics[width=\linewidth]{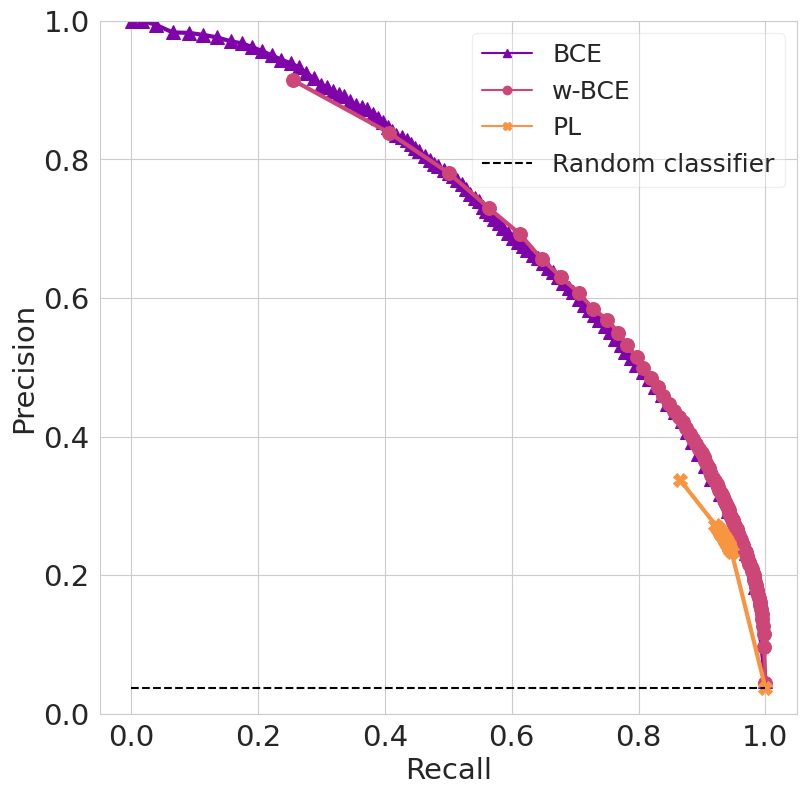}
    \label{fig:LossandPCimpact_d}
\end{subfigure}
\caption{Impact of loss function and critical probability on learning performance for NN-hybrid trained on the Mediterranean region, for a window size of 3 h, with the $95^{th}$ quantile as rainfall threshold. (a) Predicted probabilities density by loss function. (b) TP, FP and FN values in percentage by loss function. In green: True Positives (TP), in orange: False Positives (FP), in red: False Negatives (FN). (c) Critical probability impact on the PSS for each loss function. (d) Precision-Recall curve for each loss function. In purple: Binary Cross-Entropy (BCE), in pink: Weighted Binary Cross-Entropy (w-BCE), in orange: Peirce Loss (PL)}
\label{fig:LossandPCimpact}
\end{figure*}

This subsection discusses how the use of the three loss functions previously defined in Sec. 2.f---namely BCE, w-BCE and PL---affects the obtained results. In particular, the choice of the loss function significantly impacts the probability distribution of the issued probabilistic forecasts, and therefore performance once the critical probability $P_C$ is fixed.
Results are showed in Figure~\ref{fig:LossandPCimpact} for $\Delta t = 3~h$ and the $95^{th}$ percentile as rainfall threshold.

The empirical distribution of HPE outcome probability $p_i$ obtained according to the three losses are displayed on Figure \ref{fig:LossandPCimpact_a}. 
The BCE (purple bars) provides most probabilities on a range between $0$ and $0.1$, with very few predicted probabilities above $0.1$. This means that for the ''natural'' choice $P_C = 0.5$, most predicted events would not exceed the rainfall threshold.
The w-BCE (pink bars) allows to more homogeneously distribute the frequencies of predicted probabilities, thus allowing to predict many more threshold exceedances when setting the same threshold $P_C = 0.5$.
Finally, the PL (yellow bars) results in a distribution which is more binary, where the large majority of predicted probabilities are either between $0$ and $0.1$ or between $0.9$ and $1$, with no predicted probabilities in between. The consequence of such a binary behavior is that the predicted class would be the same for a large range of $P_C$, which frees up the choice of a critical probability.
The differences in behavior between the three loss functions are further illustrated in Figure \ref{fig:LossandPCimpact_b}, which displays TP, FP and FN values for $P_C = 0.5$. As previously mentioned, with the BCE, most predicted probabilities remain below $P_C$, resulting in a predominance of TN (not shown) and FN (red bars), while FP (orange bars) are extremely rare.
When using the w-BCE or PL, TP (green bars) increase, meaning that rain events are correctly identified more frequently. However, this also leads to a significant rise in FP. This outcome is expected, as assigning greater importance to the under-represented class implies that misclassifying an event as negative is considered far more costly than the opposite, aligning with the objectives of this study.

Figures \ref{fig:LossandPCimpact_c} and \ref{fig:LossandPCimpact_d} offer a more in depth view on each loss function behavior. In particular, Figure \ref{fig:LossandPCimpact_c} shows the sensitivity of the PSS to $P_C$ for each loss function, while Figure \ref{fig:LossandPCimpact_d} shows the associated Precision-Recall curve. 
As shown by the purple line in Figure \ref{fig:LossandPCimpact_c}, the BCE appears very sensitive to the choice of $P_C$, and the PSS performance  drops rapidly with increasing $P_C$ values. A peak in PSS is observed for $P_C^{\star} = s = 0.05$, that is the relative frequency $s$ of the class 1 event ($Y=1$). This value corresponds to the theoretical value which optimizes PSS as shown by \citet{Mason1979}. 
Most values for the BCE are then located at high Precision and low Recall (purple symbols on Figure \ref{fig:LossandPCimpact_d}). 
When using the w-BCE (pink line in Figure \ref{fig:LossandPCimpact_c}), the behavior of PSS as a function of $P_C$ is maximum around $P_C^\star \approx 0.5$ and is more stable over a wide range of $P_C$ (between $0.2$ and $0.8$). This observation is explained by simple arguments in Appendix~A.
Most corresponding values in Figure \ref{fig:LossandPCimpact_d} are found for high Recall  and low Precision.
Finally, when using the PL (orange line in Figure \ref{fig:LossandPCimpact_c}), the PSS performance is almost constant for all values of $P_C$, with the exception of extreme values close to $0$ or $1$. This stability is reflected in the Precision-Recall curve (Figure \ref{fig:LossandPCimpact_d}), where a cluster of values is located at high Recall and low Precision.
 
In summary, as illustrated by the empirical results of Fig \ref{fig:LossandPCimpact} and explained in Appendix~A, both balanced loss functions (w-BCE and PL) provide a greater stability with $P_C$, making the standard choice of $P_C=0.5$ relevant for PSS. The BCE shows sensitivity to the choice of the threshold probability $P_C$ and requires an optimized selection of a $P_C$ value to match PSS and Recall performances of the two balanced loss function.
%
%
%
%
%
%
\subsection{NN-Hybrid performance for HPE prediction}
\begin{figure*}[h!]
    \centering
    \begin{subfigure}{0.35\linewidth}
        \caption{}
        \includegraphics[height=5.cm,trim={0.2cm 0.2cm .5cm 0.2cm},clip]{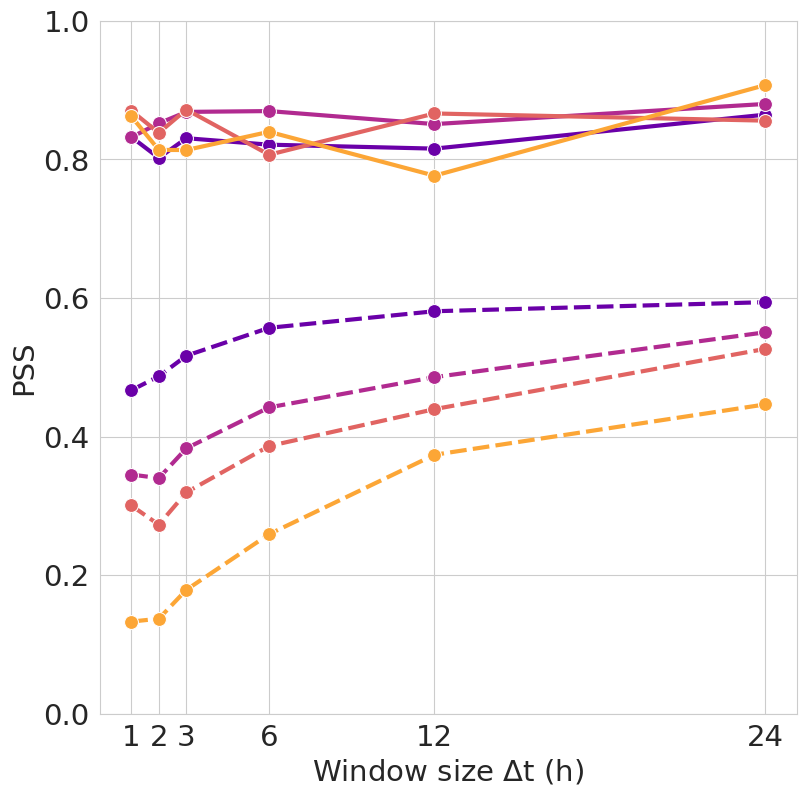}
        \label{fig:HPE_BCE}
    \end{subfigure}%
    \begin{subfigure}{0.3\linewidth}
        \caption{}
        \includegraphics[height=5.cm,trim={2.5cm 0.2cm .5cm 0.2cm},clip]{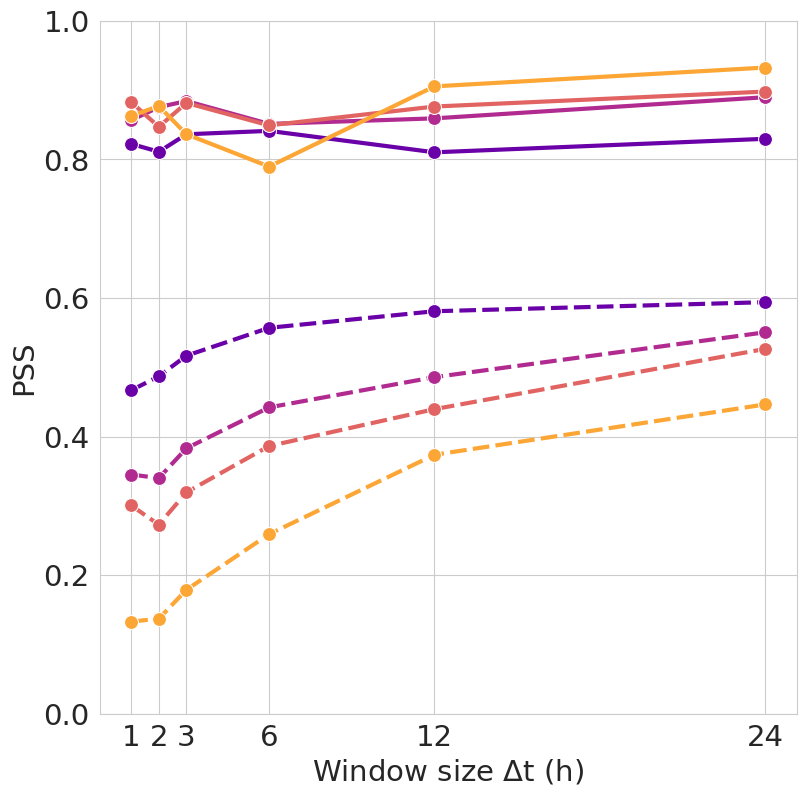}
        \label{fig:HPE_wBCE}
    \end{subfigure}%
    \begin{subfigure}{0.3\linewidth}
        \caption{}
        \includegraphics[height=5.cm,trim={2.5cm 0.2cm .5cm 0.2cm},clip]{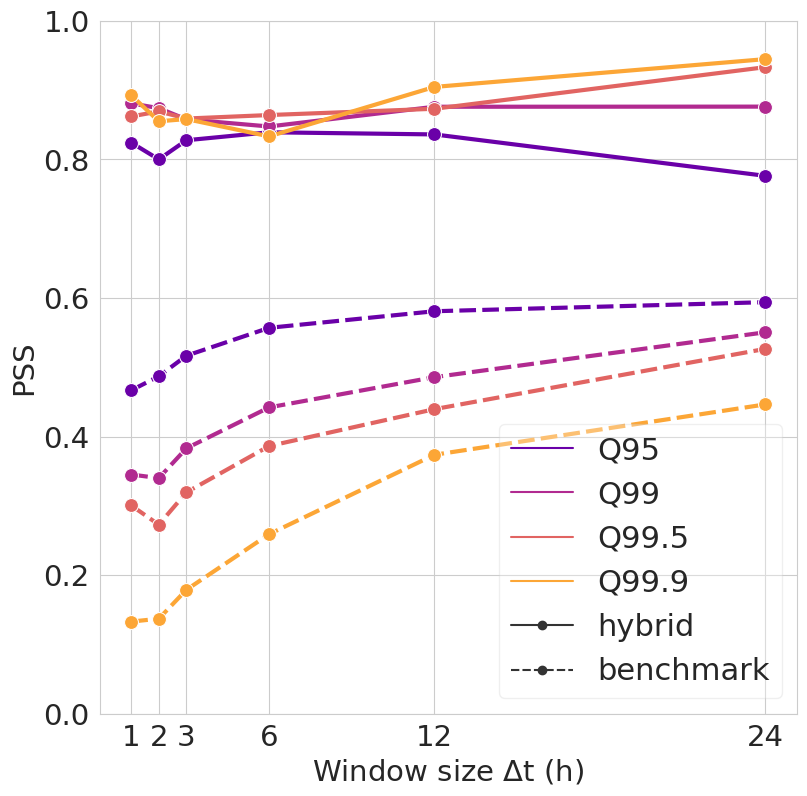}
        \label{fig:HPE_PL}
    \end{subfigure}
    \caption{Results for predicting increasingly intense rainfall events with NN-Hybrid trained on Mediterranean hybrid dataset, with (a) the Binary Cross-Entropy (BCE), (b) the Weighted Binary Cross-Entropy (w-BCE) and (c) the Peirce Loss (PL) as loss function. For (a), $P_C=0.05$, $P_C=0.01$, $P_C=0.005$ and $P_C=0.001$, respectively for the $95^{th}$, $99^{th}$, $99.5^{th}$ and $99.9^{th}$ quantile. For (b) and (c), $P_C = 0.5$.
    From coldest to warmest color: $95^{th}$ quantile, $99^{th}$ quantile, $99.5^{th}$ quantile and $99.9^{th}$ quantile. Solid line: NN-Hybrid predictions. Dashed and dotted line: Benchmark predictions.}
    \label{fig:HPE}
\end{figure*}

Figures \ref{fig:HPE_BCE}, \ref{fig:HPE_wBCE} and \ref{fig:HPE_PL} summarize the performance of NN-Hybrid compared with the benchmark for increasingly intense rainfall events and different loss functions, respectively the BCE, the w-BCE and PL.

For all rainfall thresholds, the benchmark (in dashed lines) displays results that are consistent with what was discussed previously in subsection \ref{sub_BHD} and showcased on Figure \ref{fig:PSSvsdeltat}, namely better performances for longer $\Delta t$ than shorter ones. However, performances decrease with increasing rainfall threshold (from purple to pink, orange and yellow). In other words, the more intense and rare the event to predict, the more the benchmark struggles to correctly predict its occurrence.
Figure \ref{fig:HPE_BCE} shows that, when using the BCE with $P_C$ set equal to the event probability frequency $s$, the PSS performances of NN-Hybrid (solid lines) is significantly better than the benchmark, reaching PSS values near $0.8$, with little sensitivity to rainfall thresholds. 
Figures \ref{fig:HPE_wBCE} and \ref{fig:HPE_PL} show similar results when using a balanced loss function with a standard choice of $P_C=0.5$. First, with both the w-BCE and the PL, all NN-Hybrid predictions are reaching PSS values above $0.7$, outperforming the benchmark for all $\Delta t$ and rainfall threshold values. Moreover, the PSS performances of NN-hybrid with both balanced loss function are stable with increasing rainfall threshold.

Summing up, the results reported on Figure \ref{fig:HPE} show that the NN-Hybrid model is suitable for the prediction of HPE for all the considered quantiles. Indeed, the three methods show stable performances with increasing rainfall threshold, which means that NN-hybrid can predict the occurrence of the rarest 5\% of HPE as well as the 0.1\%. These results also confirm that performances are similar for the three different tested loss functions, provided that $P_C$ has been chosen in a relevant manner.

\subsection{Difference in predictability between regions and case studies}
\begin{figure*}[h!]
    \centering
    \begin{subfigure}{0.45\linewidth}
        \centering
        \caption{}
        \includegraphics[width=0.9\linewidth]{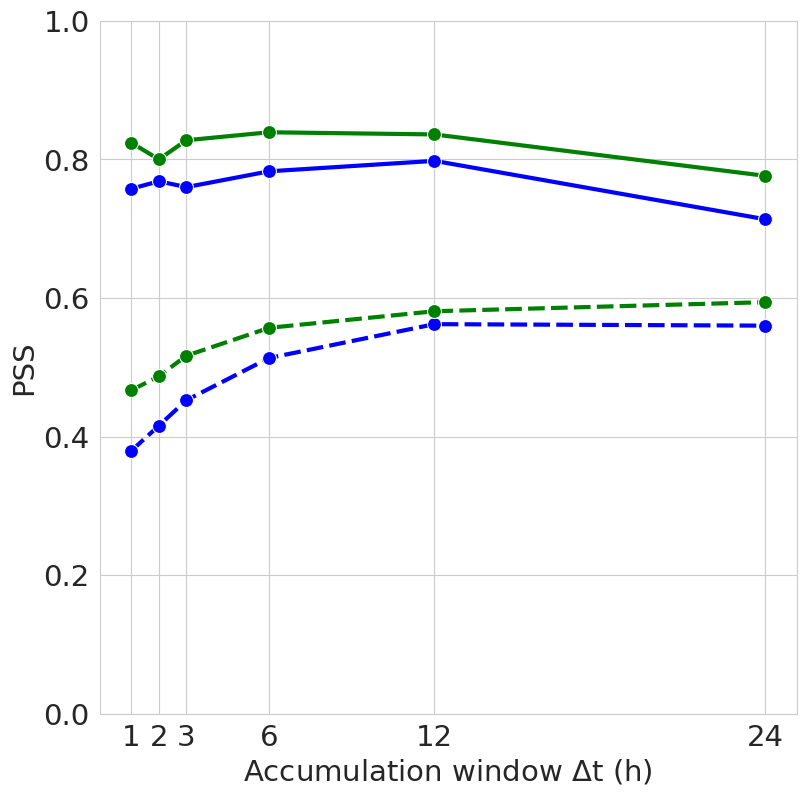}
        \label{fig:geography}
    \end{subfigure}
    \begin{subfigure}{0.45\linewidth}
        \centering
        \caption{}
        \includegraphics[width=0.9\linewidth]{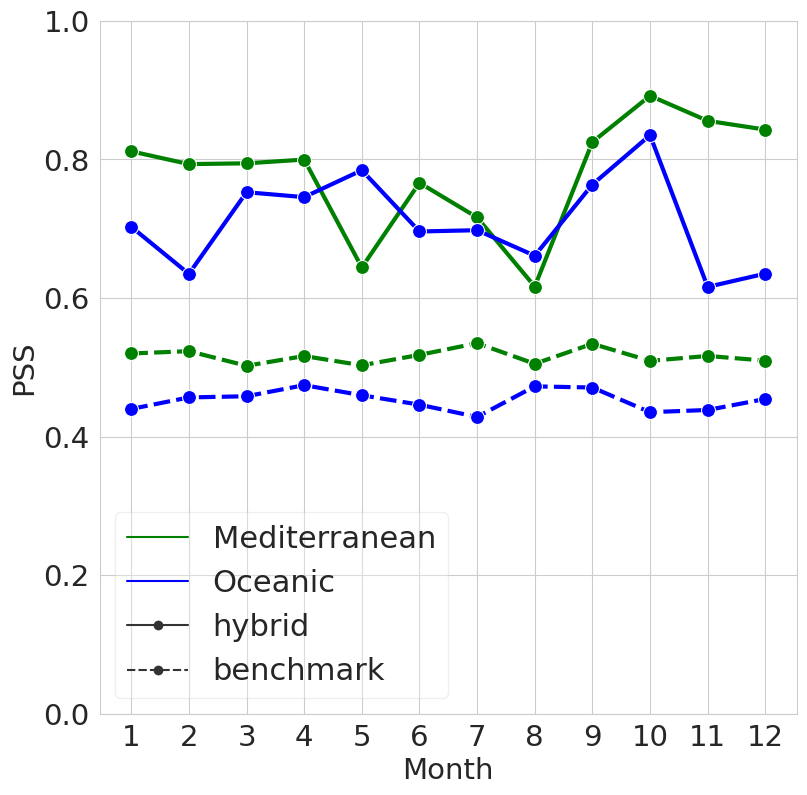}
        \label{fig:season}
    \end{subfigure}
    \caption{Regional and seasonal impacts on the PSS performance. (a) Sensitivity to $\Delta t$ and (b) sensitivity to the month of year, for the oceanic (blue) and Mediterranean (green) regions, and NN-Hybrid (solid lines) and Benchmark predictions (dashed lines). In both panels, $P_C = 0.5$ and the rainfall threshold is the $95^{th}$ quantile, with the Peirce Loss as loss function.}
    \label{fig:geography_and_season}
\end{figure*}

Figure \ref{fig:geography_and_season} highlights the PSS performance for the NN-Hybrid model by region and month of the year, with the $95^{th}$ quantile as rainfall threshold, $P_C = 0.5$ and the PL as loss function.
Figure \ref{fig:geography} first shows the sensitivity to $\Delta t$ for the oceanic region (in blue) and the Mediterranean (in green). These results reveal consistently better performance in the Mediterranean region, for all $\Delta t$, for both NN-Hybrid and benchmark predictions.
Moreover, Figure \ref{fig:season} provides further detail on the two regions by comparing PSS performance by month of the year. While benchmark performances are stable with seasons for the two regions, a drop in performance on the PSS for NN-Hybrid predictions is seen during the summer in the Mediterranean, performances being better during fall and winter. In the oceanic region, however, no clear seasonal cycle is observed. This can be explained by the dry summers of the Mediterranean climate, where few days of rain are observed, while the oceanic climate is prone to more frequent rainy days as seen in Figure \ref{fig:rainy_day}. It is noteworthy that the performance on the PSS in the Mediterranean peaks during the autumn season, which is when most HPEs occur in this region. 

\begin{figure*}[h!]
\centering
\begin{subfigure}{.41\textwidth}
  \centering
    \caption{}
  \includegraphics[width=\linewidth,trim={0.2cm 3.cm .5cm 0.2cm},clip]{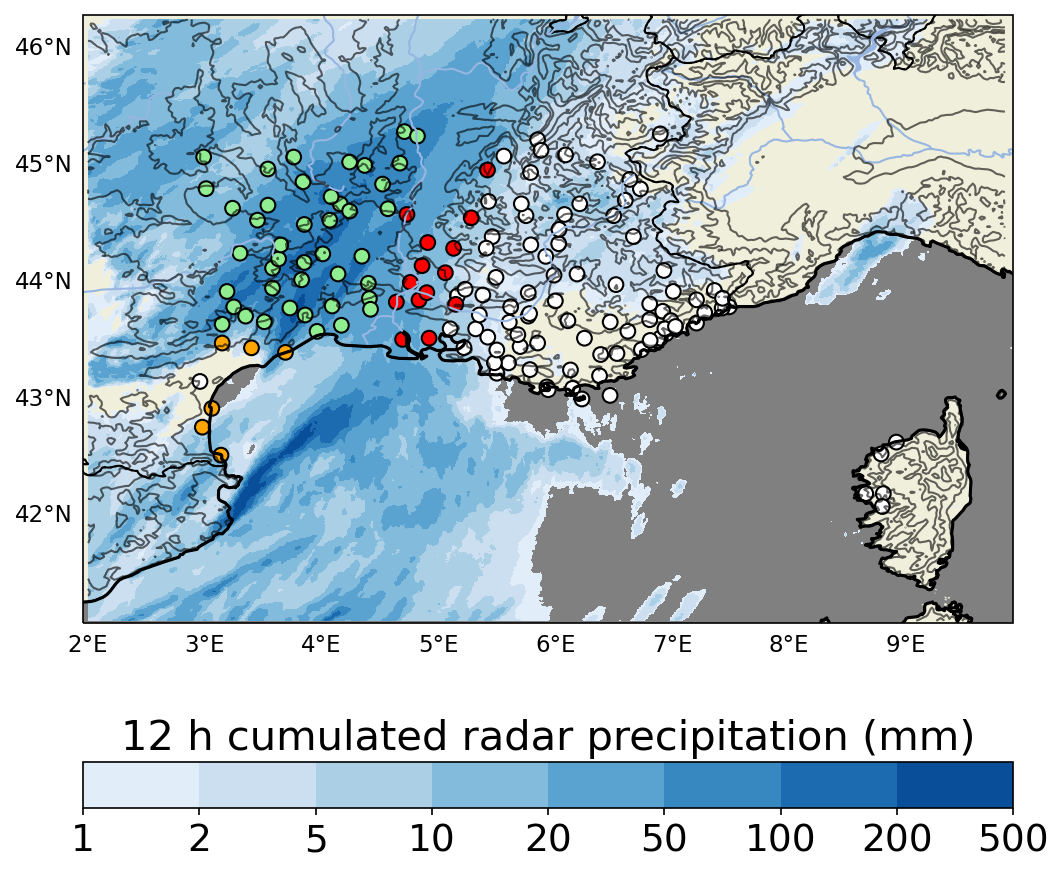}
  \label{fig:MedCase_benchmark}
\end{subfigure}
\begin{subfigure}{.41\textwidth}
  \centering
  \caption{}
  \includegraphics[width=\linewidth,trim={0.2cm 3.cm .5cm 0.2cm},clip]{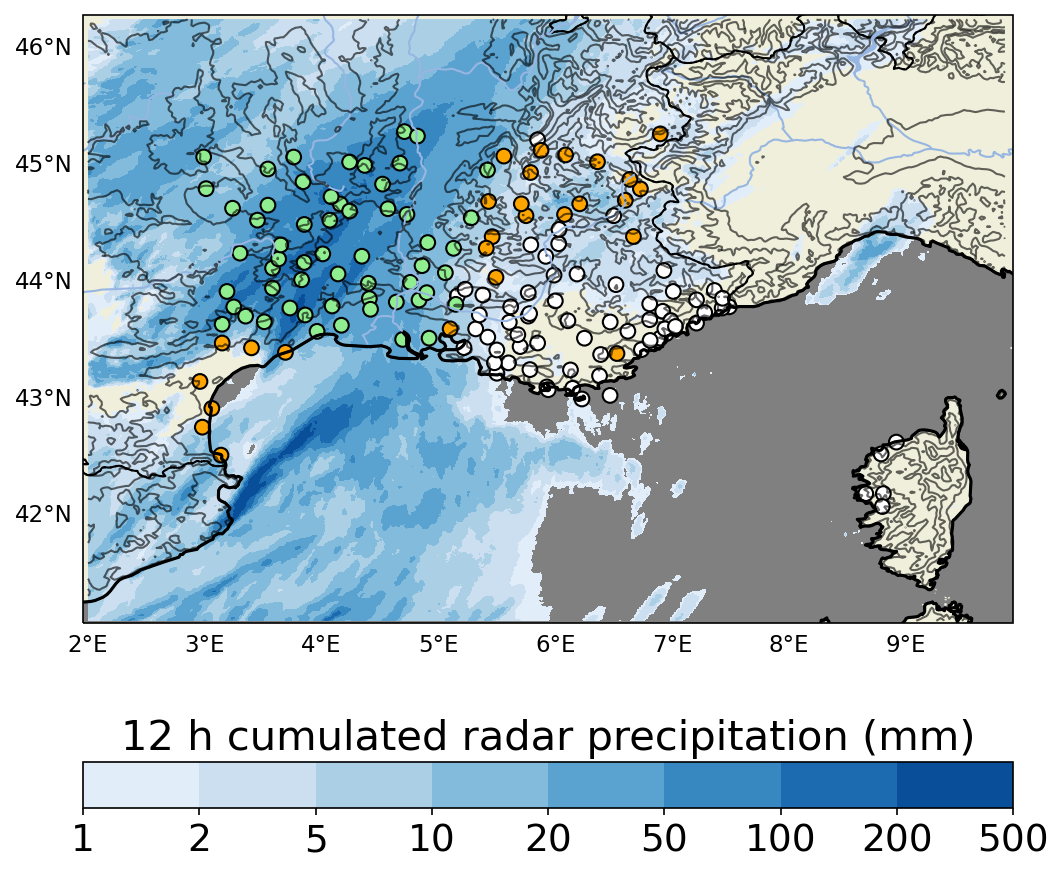}
  \label{fig:MedCase}
\end{subfigure}

\begin{subfigure}{0.41\linewidth}
    \centering
    \caption{}
    \includegraphics[width=\linewidth,trim={0.2cm 3.cm .5cm 0.2cm},clip]{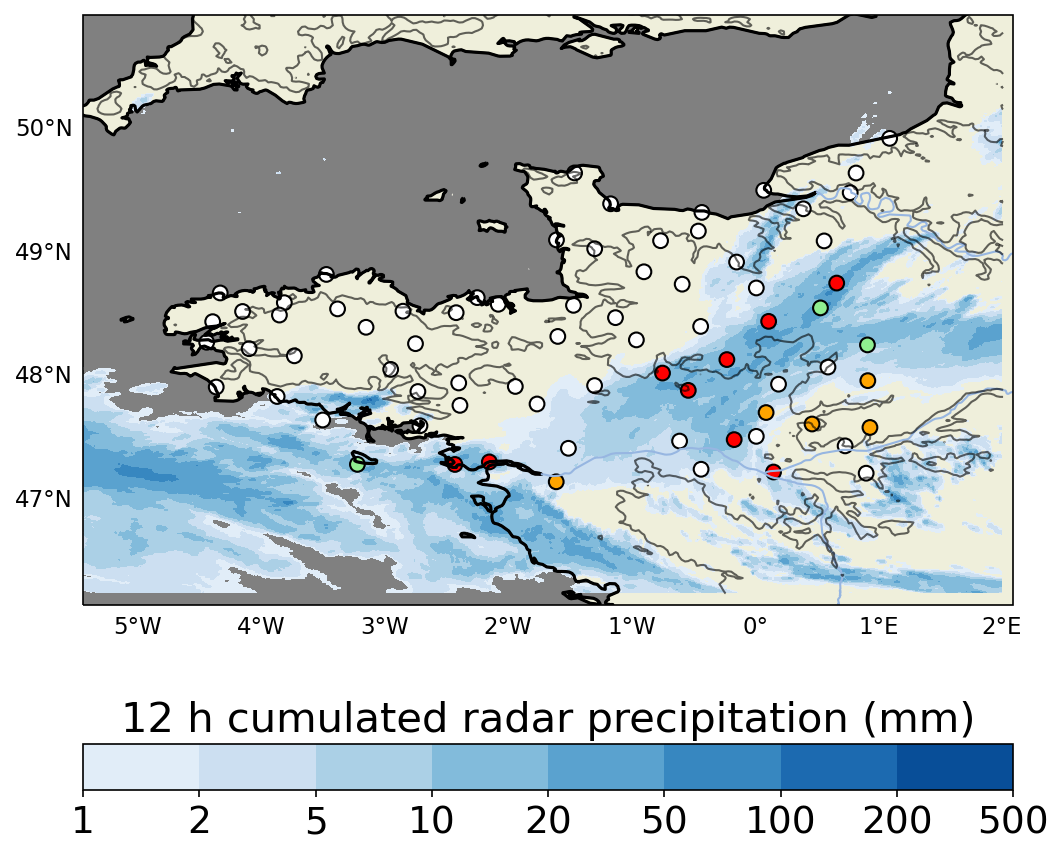}
    \label{fig:OceCase_benchmark}
\end{subfigure}
\begin{subfigure}{0.41\linewidth}
    \centering
    \caption{}
    \includegraphics[width=\linewidth,trim={0.2cm 3.cm .5cm 0.2cm},clip]{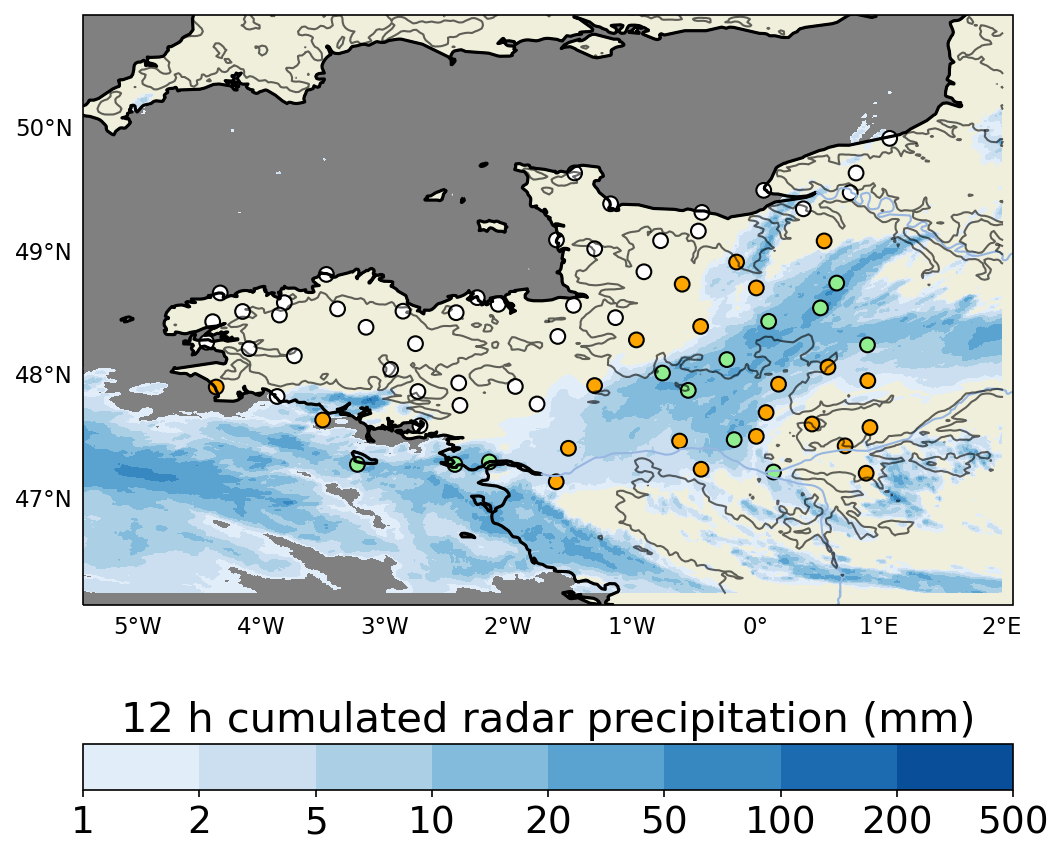}
    \label{fig:OceCase}
\end{subfigure}
\begin{subfigure}{0.41\linewidth}
    \centering
    \includegraphics[width=\linewidth,trim={0.cm 0.cm 0.cm 11.cm},clip]{20160529_0_hybrid_Q95.png}
    \label{fig:OceCase}
\end{subfigure}
\caption{Example case studies of rainfall threshold exceedance forecast: 12~h cumulated precipitation prediction from Arome ((a) and (c)) and NN-Hybrid ((b) and (d)) on 4 November 2017 from 12~UTC to 24~UTC in the Mediterranean region ((a) and (b)) and on 29 May 2016 from 00~UTC to 12~UTC in the oceanic region ((c) and (d)). White, green, orange and red dots represent TN, TP, FP and FN, respectively, with the $95^{th}$ quantile as rainfall threshold, the Weighted Binary Cross-Entropy as loss function and $P_C = 0.5$. The color shading represents 12~h cumulated precipitation observed by radar and gray contours represent altitude.}
\label{fig:Cases}
\end{figure*}

Instances on HPE prediction are shown in figures \ref{fig:Cases} for both the oceanic and Mediterranean regions to illustrate the NN-Hybrid forecast compared to the raw Arome forecast.

The Mediterranean example (panels a--b) is an event with southeasterly moist air-flow coming from the Mediterranean impinging on the Massif Central orography, leading to intense convection and thunderstorms. The maximum cumulated precipitation observed between 12~UTC and 24~UTC reaches 167.1~mm from ground stations and 340~mm over the sea from radars. The oceanic example (panels c--d) is associated with a surface low-pressure area located over Germany that forced a southwesterly flow coming from the Atlantic. The maximum cumulated precipitation observed between 00~UTC and 12~UTC reaches comparatively lower values of 51.9~mm from ground stations and 100~mm over the sea from radars.
The forecast improvement of NN-Hybrid (panels b, d) compared to Arome (panels a, c) can be seen for both events. For a significant number of stations, Arome fails to predict rainfall threshold exceedance, as shown by many FN (red points) observed in locations where precipitation were intense. On the other hand, NN-Hybrid successfully forecasts rainfall threshold exceedance and TP (in green) are observed where Arome fails whereas no FN is seen. However, as expected, the number of FP (in orange) is increased with NN-Hybrid. TN (in white) are generally well predicted by both Arome and NN-Hybrid, although Arome performs better, again as expected.
Thus, the case studies illustrate the improvement brought by NN-hybrid in both climates for not missing actual and potentially high-impact events but at the price of more frequent false alarms.
%
%
%
%
%
%
\section{Conclusions}
\label{sec:Conclusions}


Heavy Precipitation Events (HPE) in the Mediterranean can be the cause of great damage and casualties, therefore, it is necessary to anticipate their occurrence.
However, despite considerable improvement in the past decades, numerical weather prediction (NWP) models still struggle with forecasting Mediterranean HPE \citep{Khodayar2021}.
This work proposes an approach to improve precipitation forecast with neural networks (NN) using data from both NWP and station observations. It focuses on high rainfall threshold exceedance at specific locations for time windows from 1 to 24~h. The Peirce Skill Score is used for forecast verification, as it is an equitable metric suited for rare event predictions.

Results show that NN forecast using data from the global Arpege and regional Arome NWP considerably improve precipitation forecast compared to raw Arome NWP. When using only observed data the NN offers similar performances but for short time windows only. 
This is consistent with literature, as most studies which leverage only observed data face a rapid drop in performance with increasing lead times \citep[e.g.][]{andrychowicz2023}. 

The hybrid dataset---NN-Hybrid, combining observed and NWP data---offers the best performances overall. By combining the best of both types of data, a clear stability with time window is found. An additional benefit of NN-Hybrid is that rainfall threshold exceedance is equally well predicted when considering high thresholds between the $95^{th}$ and the $99.9^{th}$ quantiles, for each region and each time window. Results also show statistically better performances on Mediterranean than oceanic climate, with a peak performance during fall and winter, seasons during which most Mediterannean HPE occur \citep{Ricard2012}.
The approach of this study is similar to MetNet2 Hybrid from \citet{espeholt2022}, where the authors present a certain stability with increasing rainfall threshold for short lead times, although their performances drop for highest rainfall threshold and lead time greater than 1~h.

These results depend on the informed choice of the loss function and the critical probability $P_C$ associated, which is guided by the trade-off between missing as few HPE as possible and avoiding over-predicting the occurrence of HPE. Three loss functions are compared: the classical Binary Cross-Entropy, and both balanced Weighted Binary Cross-Entropy and Peirce Loss introduced here. The two balanced loss functions portray a binary behavior, thus results are stable with regard to the choice of $P_C$. On the other hand, the Binary Cross-Entropy outputs predicted probabilities close to the actual frequencies of events to be predicted, which requires to carefully chose $P_C$. By adjusting the value of $P_C$ to the corresponding probability of the event (e.g. $P_C = 0.05$ for the $95^{th}$ quantile), similar results are found with the Binary Cross-Entropy than with the two balanced loss functions with a standard choice of $P_C = 0.5$. Applying this configuration with the three loss functions results in a considerable decrease in false negative rate with the downside of having an increased false positive rate. Given the severity of HPE, this trade-off is considered acceptable and is consistent with the scope of the paper.


It is important to mention the limitations of this study. 
First, the developed NN-models only provide forecast for different rainfall accumulation window sizes, but the sensitivity with delays from NN-model initialization is not investigated. Additionally, the available NWP is always initialized at 00~UTC, which influences lead time sensitivity. A more detailed study is needed to see whether the stability of NN-models (in particular NN-Hybrid) is still valid with delays from initialization to better match the framework used in operational forecasts and in other studies \citep{espeholt2022}.

Additionally, the period covered by the database (2016 to 2018) is relatively short, as is the number of HPE to learn from. Having a longer database would allow to learn and validate from more extreme cases, which also are the most destructive ones. However, NWP data are non-homogeneous as NWP models are improved regularly, and high-resolution NWP models such as Arome only cover one or two decades at best. Using reforecast data would allow to alleviate the former limitation by offering homogeneous data over a long period and an up-to-date model version \citep[e.g.][]{Doiteau2024}. Moreover, our work focuses exclusively on deterministic forecasts, while a probabilistic approach could also be highly relevant for predicting Mediterranean HPEs \citep{Grazzini2024}. Exploring such probabilistic methods will be an important direction for future research. We also remark that using NWP data with higher resolution may be particularly useful, especially considering the steep mountains of the Mediterranean, as a higher resolution would enable convection to be fully represented and the impact of orography to be more accurate.

On another note, even though the goal of this study is to compare different approaches rather than to propose an innovative NN architecture, we acknowledge that the NN models used in this study, i.e. a combination of recurrent and convolutional NN, are somewhat outdated compared to the latest developments in the field. Using more recent architecture such as transformers or Generative Adversarial Networks could help to better extract spatial and temporal information from features data. However, these methods are more computationally expensive than the NN used here. Finally, having an explainable NN-model would allow to better understand how the model works, which could help meteorological research in understanding intense weather events \citep{Mcgovern2019}.

\acknowledgments
This work was supported by ANR, France grant SAPHIR project ANR-21-CE04-0014-03.

%
%
\datastatement
The data used in this work come from the MeteoNet database \citep{Meteonet}, which is publicly available and can be found at https://meteonet.umr-cnrm.fr/dataset/

\appendix[A]
\appendixtitle{Optimal threshold value and threshold sensitivity for PSS}
\label{App:opt}
\begin{figure*}[h!]
    \begin{centering}
        \includegraphics[width=0.8\linewidth]{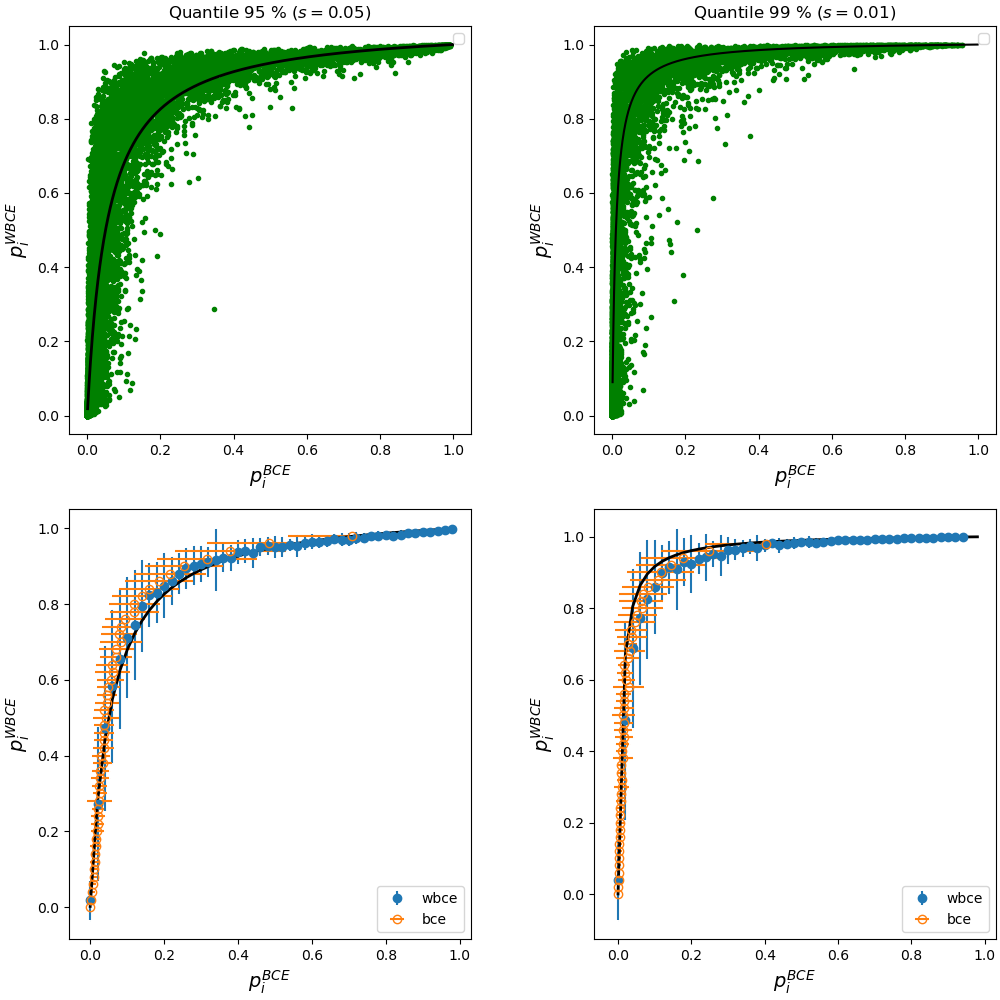}
    \end{centering}
    \caption{Relationship between probabilistic predictions based on BCE ($p_i^{\scriptscriptstyle BCE}$) and w-BCE ($p_i^{\scriptscriptstyle w-BCE}$) losses. On top panels are displayed, for each elements of validation and test samples, the obtained values of $p_i^{\scriptscriptstyle w-BCE}$ vs $p_i^{\scriptscriptstyle BCE}$ in the prediction of occurrence of threshold exceedance of the cumulative precipitation for $\Delta t = 3h$ in the case of Q95 (top left panel) and Q99 (top right panel) quantiles. In the bottom panels are plotted the corresponding conditional empirical mean and rms estimated either when $p_i^{\scriptscriptstyle BCE}$ is fixed (blue ($\bullet$) symbols) or when $p_i^{\scriptscriptstyle w-BCE}$ is fixed (orange ($\circ$) symbols). In all panels, the solid black lines represent the expected relationship as respect to Eq. \eqref{eq:pwbce}}. 
    \label{fig:check_wbce}
\end{figure*}

In this Appendix, we examine the relationship between \( q_i \), the probability that \( Y_i = 1 \), and \( p_i \), the output of a model trained to minimize each of the three losses defined in Sec. \ref{sec:Data}.\ref{subsec:loss}. This problem is closely related to the calibration of the predicted distribution, a key aspect in probabilistic forecasting.

In practice, the exact relationship between \( q_i \) and \( p_i \) is difficult to determine, as \( q_i \) cannot be observed. Here, we consider an idealized scenario in which each \( q_i \) is assumed to be perfectly known. Under this assumption, we can replace \( Y_i \) with its probabilistic counterpart \( q_i \) in Eqs. \eqref{eq:bce}, \eqref{eq:wbce}, and \eqref{eq:pssloss}.  
We also assume in \eqref{eq:pssloss} that $N$ is large enough so that $N^{-1} \sum_{i=1}^N Y_i \simeq s$. In that case, 
the value that minimizes the BCE (that turns out to be the negative log-likelihood) is simply:
\begin{equation}
\label{eq:pbce}
  p_i^{\scriptscriptstyle BCE} = q_i \; .
\end{equation}
According to Eq. \eqref{eq:wbce}, $p_i^{\scriptscriptstyle WBCE}$ that minimizes the w-BCE loss has to satisfy
$\frac{q_i}{s p_i}- \frac{(1-q_i)}{(1-s)(1-p_i)} = 0$, which leads to:
\begin{equation}
\label{eq:pwbce}
 p_i^{\scriptscriptstyle w-BCE} = \frac{q_i(1-s)}{q_i(1-s)+s(1-q_i)} \; .
\end{equation}

Let us notice that if $s=1/2$, w-BCE reduces to the BCE, as expected, we recover 
$p_i^{\scriptscriptstyle w-BCE} = p_i^{\scriptscriptstyle BCE}$.
Finally, it is straightforward to see that PL is optimized by:
\begin{equation}
\label{eq:ppl}
  p_i^{\scriptscriptstyle PL} = 
  \begin{cases}
1 & \text{if} \; \; q_i \geq s \\
0 & \text{otherwise}
\end{cases}
\end{equation}
The previous equation means that the PL output is directly $Y_i$ with $P_C = s$.

Even if we do not have access to $q_i$,  empirically we can check that equations \eqref{eq:pwbce} and \eqref{eq:ppl} are meaningful since, thanks to \eqref{eq:pbce}, they entail 
a specific relationship of $p_i^{\scriptscriptstyle w-BCE}$ and $p_i^{\scriptscriptstyle PL}$ as functions of $p_i^{\scriptscriptstyle BCE}$. This is illustrated in Fig. \ref{fig:check_wbce} 
using prediction data of the hybrid model for $Q95$ and $Q99$ exceedance of $\Delta t = 3$ h cumulative precipitation in SE ground stations.

Eqs. \eqref{eq:pbce}, \eqref{eq:pwbce}, and \eqref{eq:ppl} are also useful for determining the optimal threshold \( P_C^\star \) for a given score. In particular, since the optimal threshold for the PSS score is given by \( q^\star = s \), as shown by \citet{Mason1979} (see also \citet{Jolliffe2004}), a value that therefore corresponds to \( P_C^\star = s \) in the case of the BCE loss.  It results, from \eqref{eq:pwbce}, that the optimal value for w-BCE becomes \( P_C^\star = \frac{1}{2} \), independent of \( s \). This latter value is close to the empirically observed optimal threshold in Fig. \ref{fig:LossandPCimpact}(c). For PL, as it can be seen from \eqref{eq:ppl}, there is no well defined $P_C^\star$ and regardless of the chosen \( P_C \) (e.g., \( P_C = \frac{1}{2} \)), the predicted class \( Y_i \) remains unchanged and so the PSS value. This is also what we observe in Fig. \ref{fig:LossandPCimpact}(c). Note that if the optimal threshold for a given metric is \( q^\star = \frac{1}{2} \), then the best threshold for w-BCE is \( P_C^\star = 1 - s \), whereas PL is not well-suited for this score.

Eqs. \eqref{eq:pbce}, \eqref{eq:pwbce}, and \eqref{eq:ppl} can be further leveraged to compare the behavior of the score \( S(q) \) around its optimal value \( q = q^\star \) when using different loss predictions. Specifically, if \( S(q) \) corresponds to \( S(p^{\scriptscriptstyle BCE}) \), then the score computed for w-BCE, denoted as \( S_1(p^{\scriptscriptstyle WBCE}) \), is given by  
\[
S_1(p^{\scriptscriptstyle w-BCE}) = S\Big(q(p^{\scriptscriptstyle w-BCE})\Big),
\]
where \( q(p) \) represents the inverse function of expression \eqref{eq:pwbce}. A straightforward algebraic derivation shows that at \( q = q^\star \), where the first-order derivative of \( S(q) \) vanishes, the ratio of the second-order derivatives of \( S \) and \( S_1 \) reads:
\begin{equation}
\label{eq:ratio_d2}
  r^\star = \frac{S_1''(P_C^\star)}{ S_0''(q^\star)} = \frac{s(1-s)}{\Big(1 - 2P_C^{\star}(1-s) - P_C{^\star}\Big)^2}.
\end{equation}
In the case of the PSS, where \( P_C^\star = 1/2 \), this ratio takes values close to \( r^\star = 1/5 \) and \( r^\star = 1/25 \) for the \( Q95 \) and \( Q99 \) thresholds, respectively. As observed in Fig. \ref{fig:LossandPCimpact}(c), for high quantile thresholds, the PSS score based on w-BCE exhibits significantly lower sensitivity around its maximum than the one obtained with BCE, making it more stable over a wider range of \( P_C \) values.  

This effect is even more pronounced when using the PL loss, for which the second derivative theoretically vanishes. Consequently, as emphasized earlier, the PSS score remains independent of the choice of \( P_C \).

\appendix[B]
\label{App:HSS_CSI}
\appendixtitle{Alternative classification scores}

\begin{figure*}[h!]
    \centering
    \begin{subfigure}{0.45\linewidth}
        \centering
        \caption{}
        \includegraphics[width=0.9\linewidth]{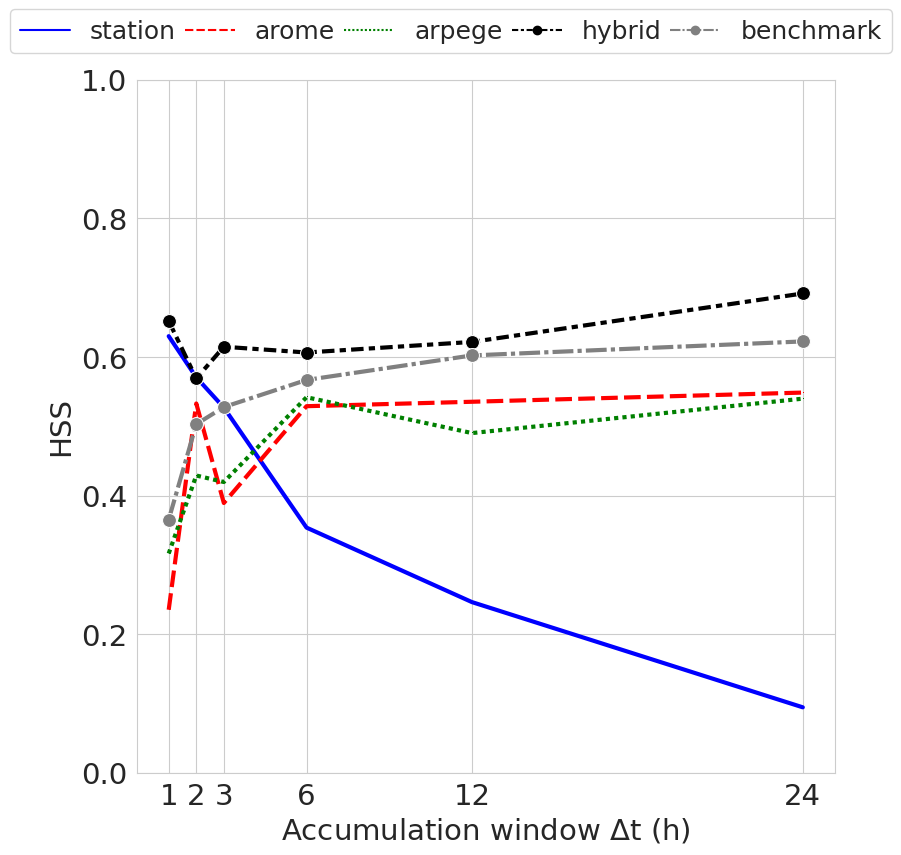}
        \label{fig:HSS}
    \end{subfigure}
    \begin{subfigure}{0.45\linewidth}
        \centering
        \caption{}
        \includegraphics[width=0.9\linewidth]{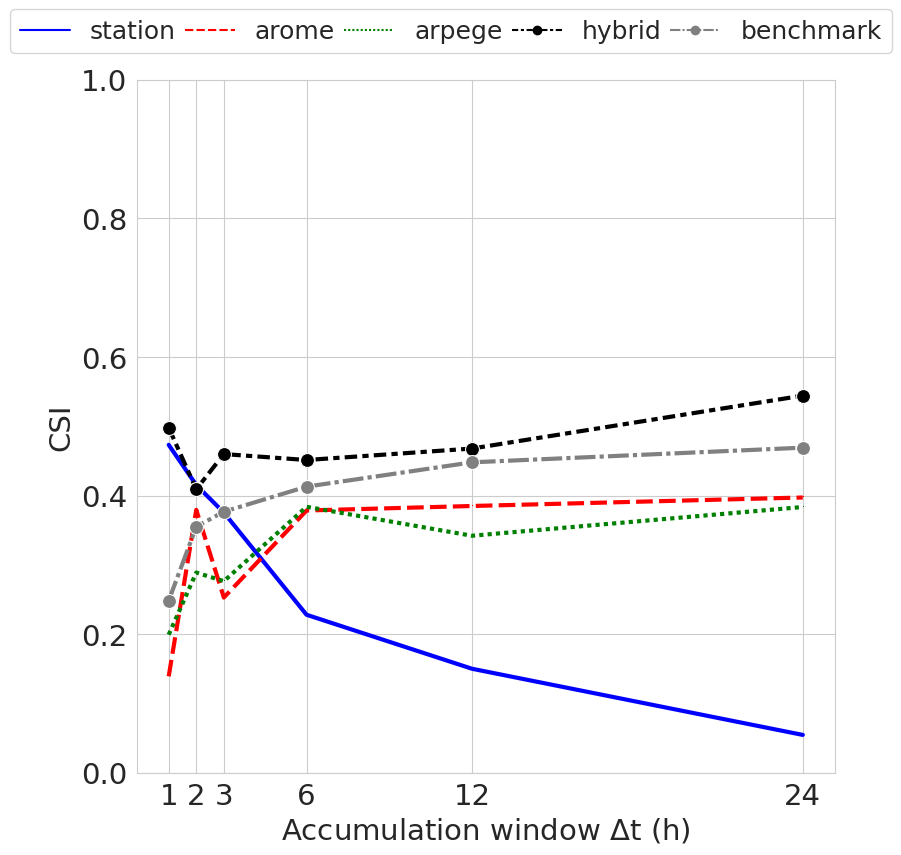}
        \label{fig:CSI}
    \end{subfigure}
    \caption{HSS (a) and CSI (b) scores as the function of the window size $\Delta t$ for various NN-models and the benchmark. The HPE considered are the 5\% ($s = 0.05$) rarest rainfall events ($95^{th}$ quantile) and the loss is the Binary Cross-Entropy loss with $P_C = 0.5$.  In solid blue: NN-Station. In dashed red: NN-Arome. In dotted black: NN-Hybrid. In dotted and dashed gray: Benchmark performance. Even if $P_C = 0.5$ can be suboptimal for both metrics, we see that the hybrid model provide the best results at all $\Delta t$.}
    \label{fig:HSS_CSI}
\end{figure*}

Besides the PSS used all along this paper, 
they are many scores that can be defined from the contingency matrix and that are commonly computed to assess the performance of binary classification in various domains \citep{Jolliffe2004,WILKS2019369}.

The {\em Heidke Skill Score} (HSS) is a widely used score that evaluates the performance of a binary forecast by measuring its accuracy as compared
to ``pure chance''. It is defined as:
\[
HSS = \frac{\frac{TP + TN}{N} - E}{1 - E}
\]
where \( \frac{TP + TN}{N} \) is the rate of correct prediction (either negative or positive) of the method to be evaluated and E is the expected rate of correct prediction obtained by chance. If
\( FP \) and \( FN \) stand for the number of false positives and false negatives forecasts, 
\(E\), the probability of correctly predicting 0 or 1, is simply the sum of the product of probabilities to predicting one class (0 or 1):
\[
E = \frac{(TP + FN)(TP + FP) + (TN + FP)(TN + FN)}{N^2} \; .
\]

The HSS yields a score ranging from \(-1 \) to \(1\). A perfect forecast achieves an HSS of \(1\), indicating perfect agreement between forecasts and observations beyond what is expected by chance. A score of \(0\) suggests that the forecast has no skill compared to the reference forecast (i.e., it is no better than chance or climatology) while negative HSS values imply that the forecast is actually worse than the reference forecast. HSS is valuable for assessing overall forecast skill but it is sensitive to class imbalance and may not be appropriate for HPE prediction.
As advocated in \cite{Jolliffe2004}, the optimal threshold probability for HSS depends on HSS itself and of $s$, the rate of observed events but when the latter is small enough, one can choose the value $P_C = 0.5$.

The {\em Critical Success Index} (CSI), also referred to as the Threat Score (TS), is another metric that can be particularly useful in the case of imbalanced datasets. 
It measures the accuracy of positive predictions while ignoring true negatives, making it useful for assessing rare event forecasting, such as HPE. Unlike PSS or HSS, CSI does not involve all the four elements of the confusion matrix. It is defined as:
\begin{equation}
  CSI = \frac{TP}{TP+FP+FN}
\end{equation}
Notice that the CSI ranges between 0, the worst value, and 1, the best possible score.  As for HSS, the optimal threshold value depends on the CSI itself so it has to be estimated empirically. However, when $s$ is small, $P_C = 0.5$ is in general not far from the optimum. 

The performance of all NN models optimized using the Binary Cross-Entropy loss, compared to the benchmark in terms of HSS and CSI scores, is reported in Fig. \ref{fig:HSS_CSI}. In both cases, we used a threshold probability \( P_C = 0.5 \), as it is expected to be close to the optimal value. 
We observe that, even for these scores, the hybrid model outperforms the other variants. Moreover, even if \( P_C \) is suboptimal, it still performs better than the benchmark across all window sizes.

%






%





\begin{thebibliography}{50}
\providecommand{\natexlab}[1]{#1}
\providecommand{\url}[1]{\texttt{#1}}
\renewcommand{\UrlFont}{\rmfamily}
\providecommand{\urlprefix}{URL }
\expandafter\ifx\csname urlstyle\endcsname\relax
  \providecommand{\doi}[1]{https://doi.org/\discretionary{}{}{}#1}\else
  \providecommand{\doi}{https://doi.org/\discretionary{}{}{}\begingroup \urlstyle{rm}\Url}\fi
\providecommand{\eprint}[2][]{\url{#2}}

\bibitem[{Agrawal et~al.(2019)Agrawal, Barrington, Bromberg, Burge, Gazen,, and Hickey}]{agrawal2019}
Agrawal, S., L.~Barrington, C.~Bromberg, J.~Burge, C.~Gazen, and J.~Hickey, 2019: Machine {Learning} for {Precipitation} {Nowcasting} from {Radar} {Images}. arXiv, \urlprefix\url{http://arxiv.org/abs/1912.12132}, arXiv:1912.12132 [cs, stat].

\bibitem[{Andrychowicz et~al.(2023)Andrychowicz, Espeholt, Li, Merchant, Merose, Zyda, Agrawal,, and Kalchbrenner}]{andrychowicz2023}
Andrychowicz, M., L.~Espeholt, D.~Li, S.~Merchant, A.~Merose, F.~Zyda, S.~Agrawal, and N.~Kalchbrenner, 2023: Deep {Learning} for {Day} {Forecasts} from {Sparse} {Observations}. arXiv, \urlprefix\url{http://arxiv.org/abs/2306.06079}, arXiv:2306.06079 [physics].

\bibitem[{Argence et~al.(2009)Argence, Lambert, Richard, Pierre~Chaboureau, Philippe~Arbogast,, and Maynard}]{Argence2009}
Argence, S., D.~Lambert, E.~Richard, J.~Pierre~Chaboureau, J.~Philippe~Arbogast, and K.~Maynard, 2009: Improving the numerical prediction of a cyclone in the mediterranean by local potential vorticity modifications. \textit{Quart.\ J.\ Roy.\ Meteor.\ Soc.}, \textbf{135~(641)}, 865--879, \doi{https://doi.org/10.1002/qj.422}, \eprint{https://rmets.onlinelibrary.wiley.com/doi/pdf/10.1002/qj.422}.

\bibitem[{Argence et~al.(2006)Argence, Lambert, Richard, S\"ohne, Chaboureau, Cr\'epin,, and Arbogast}]{Argence2006}
Argence, S., D.~Lambert, E.~Richard, N.~S\"ohne, J.-P. Chaboureau, F.~Cr\'epin, and P.~Arbogast, 2006: High resolution numerical study of the algiers 2001 flash flood: sensitivity to the upper-level potential vorticity anomaly. \textit{Advances in Geosciences}, \textbf{7}, 251--257, \doi{10.5194/adgeo-7-251-2006}.

\bibitem[{Ayzel et~al.(2020)Ayzel, Scheffer,, and Heistermann}]{ayzel2020}
Ayzel, G., T.~Scheffer, and M.~Heistermann, 2020: {RainNet} v1.0: a convolutional neural network for radar-based precipitation nowcasting. \textit{Geoscientific Model Development}, \textbf{13~(6)}, 2631--2644, \doi{10.5194/gmd-13-2631-2020}.

\bibitem[{Baggio and Muzy(2024)Baggio, and Muzy}]{Baggio2024}
Baggio, R., and J.-F. Muzy, 2024: Improving probabilistic wind speed forecasting using {M}-{Rice} distribution and spatial data integration. \doi{10.48550/arXiv.2310.12088}.

\bibitem[{Bauer et~al.(2015)Bauer, Thorpe,, and Brunet}]{bauer2015}
Bauer, P., A.~Thorpe, and G.~Brunet, 2015: The quiet revolution of numerical weather prediction. \textit{Nature}, \textbf{525~(7567)}, 47--55, \doi{10.1038/nature14956}.

\bibitem[{Baïle and Muzy(2023)Baïle, and Muzy}]{Baile2023}
Baïle, R., and J.-F. Muzy, 2023: Leveraging data from nearby stations to improve short-term wind speed forecasts. \textit{Energy}, \textbf{263}, 125\,644, \doi{10.1016/j.energy.2022.125644}.

\bibitem[{Beck et~al.(2018)Beck, Zimmermann, McVicar, Vergopolan, Berg,, and Wood}]{koppen}
Beck, H.~E., N.~E. Zimmermann, T.~R. McVicar, N.~Vergopolan, A.~Berg, and E.~F. Wood, 2018: Present and future {Köppen}-{Geiger} climate classification maps at 1-km resolution. \textit{Scientific Data}, \textbf{5~(1)}, 180\,214, \doi{10.1038/sdata.2018.214}.

\bibitem[{Bi et~al.(2023)Bi, Xie, Zhang, Chen, Gu,, and Tian}]{Pangu}
Bi, K., L.~Xie, H.~Zhang, X.~Chen, X.~Gu, and Q.~Tian, 2023: Accurate medium-range global weather forecasting with {3D} neural networks. \textit{Nature}, \textbf{619~(7970)}, 533--538, \doi{10.1038/s41586-023-06185-3}.

\bibitem[{Bresson et~al.(2009)Bresson, Ricard,, and Ducrocq}]{Bresson2009}
Bresson, R., D.~Ricard, and V.~Ducrocq, 2009: Idealized mesoscale numerical study of {Mediterranean} heavy precipitating convective systems. \textit{Meteorology and Atmospheric Physics}, \textbf{103~(1-4)}, 45--55, \doi{10.1007/s00703-008-0338-z}.

\bibitem[{Courtier et~al.(1991)Courtier, Freydier, Geleyn, Rabier,, and Rochas}]{Arpege}
Courtier, P., C.~Freydier, J.-F. Geleyn, F.~Rabier, and M.~Rochas, 1991: The arpege project at meteo france. \textit{Seminar on Numerical Methods in Atmospheric Models, 9-13 September 1991}, \textbf{II}, 193--232.

\bibitem[{Doiteau et~al.(2024)Doiteau, Pantillon, Plu, Descamps,, and Rieutord}]{Doiteau2024}
Doiteau, B., F.~Pantillon, M.~Plu, L.~Descamps, and T.~Rieutord, 2024: Systematic evaluation of the predictability of different {Mediterranean} cyclone categories. \textit{Weather and Climate Dynamics}, \textbf{5~(4)}, 1409--1427, \doi{10.5194/wcd-5-1409-2024}.

\bibitem[{Ducrocq et~al.(2008)Ducrocq, Nuissier, Ricard, Lebeaupin,, and Thouvenin}]{Ducrocq2008}
Ducrocq, V., O.~Nuissier, D.~Ricard, C.~Lebeaupin, and T.~Thouvenin, 2008: A numerical study of three catastrophic precipitating events over southern {France}. {II}: {Mesoscale} triggering and stationarity factors. \textit{Quart.\ J.\ Roy.\ Meteor.\ Soc.}, \textbf{134~(630)}, 131--145, \doi{10.1002/qj.199}.

\bibitem[{Ducrocq et~al.(2014)}]{HyMeX2}
Ducrocq, V., and Coauthors, 2014: Hymex-sop1: The field campaign dedicated to heavy precipitation and flash flooding in the northwestern mediterranean. \textit{Bull.\ Amer.\ Meteor.\ Soc.}, \textbf{95~(7)}, 1083 -- 1100, \doi{10.1175/BAMS-D-12-00244.1}.

\bibitem[{Duffourg and Ducrocq(2011)Duffourg, and Ducrocq}]{Duffourg2011}
Duffourg, F., and V.~Ducrocq, 2011: Origin of the moisture feeding the {Heavy} {Precipitating} {Systems} over {Southeastern} {France}. \textit{Natural Hazards and Earth System Sciences}, \textbf{11~(4)}, \doi{10.5194/nhess-11-1163-2011}.

\bibitem[{Ebert and Milne(2022)Ebert, and Milne}]{Ebert2022}
Ebert, P.~A., and P.~Milne, 2022: Methodological and conceptual challenges in rare and severe event forecast verification. \textit{Natural Hazards and Earth System Sciences}, \textbf{22~(2)}, 539--557, \doi{10.5194/nhess-22-539-2022}.

\bibitem[{Espeholt et~al.(2022)}]{espeholt2022}
Espeholt, L., and Coauthors, 2022: Deep learning for twelve hour precipitation forecasts. \textit{Nature Communications}, \textbf{13~(1)}, 5145, \doi{10.1038/s41467-022-32483-x}.

\bibitem[{Flaounas et~al.(2024)}]{egusphere-2024-2809}
Flaounas, E., and Coauthors, 2024: Dynamics, predictability, impacts, and climate change considerations of the catastrophic mediterranean storm daniel (2023). \textit{EGUsphere}, \textbf{2024}, 1--29, \doi{10.5194/egusphere-2024-2809}.

\bibitem[{Frnda et~al.(2022)Frnda, Durica, Rozhon, Vojtekova, Nedoma,, and Martinek}]{frnda2022}
Frnda, J., M.~Durica, J.~Rozhon, M.~Vojtekova, J.~Nedoma, and R.~Martinek, 2022: {ECMWF} short-term prediction accuracy improvement by deep learning. \textit{Scientific Reports}, \textbf{12~(1)}, 7898, \doi{10.1038/s41598-022-11936-9}.

\bibitem[{Gandin and Murphy(1992)Gandin, and Murphy}]{gandin1992equitable}
Gandin, L.~S., and A.~H. Murphy, 1992: Equitable skill scores for categorical forecasts. \textit{Mon.\ Wea.\ Rev.}, \textbf{120~(2)}, 361--370.

\bibitem[{Goodfellow(2016)}]{Goodfellow2016}
Goodfellow, I., 2016: Deep learning. MIT press.

\bibitem[{Grazzini et~al.(2024)Grazzini, Dorrington, Grams, Craig, Magnusson,, and Vitart}]{Grazzini2024}
Grazzini, F., J.~Dorrington, C.~M. Grams, G.~C. Craig, L.~Magnusson, and F.~Vitart, 2024: Improving forecasts of precipitation extremes over northern and central {Italy} using machine learning. \textit{Quarterly Journal of the Royal Meteorological Society}, \textbf{150~(762)}, 3167--3181, \doi{10.1002/qj.4755}.

\bibitem[{Hally et~al.(2014)Hally, Richard, Fresnay,, and Lambert}]{Hally2014}
Hally, A., E.~Richard, S.~Fresnay, and D.~Lambert, 2014: Ensemble simulations with perturbed physical parametrizations: Pre-hymex case studies. \textit{Quart.\ J.\ Roy.\ Meteor.\ Soc.}, \textbf{140~(683)}, 1900--1916, \doi{https://doi.org/10.1002/qj.2257}, \eprint{https://rmets.onlinelibrary.wiley.com/doi/pdf/10.1002/qj.2257}.

\bibitem[{{Intergovernmental Panel on Climate Change}(2023)}]{IPCC}
{Intergovernmental Panel on Climate Change}, 2023: \textit{Mediterranean Region}, 2233–2272. Cambridge University Press.

\bibitem[{Jolliffe(2004)}]{Jolliffe2004}
Jolliffe, I.~T., Ed., 2004: \textit{Forecast verification: a practitioner's guide in atmospheric science}. repr ed., Wiley, Chichester.

\bibitem[{Khodayar et~al.(2021)}]{Khodayar2021}
Khodayar, S., and Coauthors, 2021: Overview towards improved understanding of the mechanisms leading to heavy precipitation in the western {Mediterranean}: lessons learned from {HyMeX}. \textit{Atmospheric Chemistry and Physics}, \textbf{21~(22)}, 17\,051--17\,078, \doi{10.5194/acp-21-17051-2021}.

\bibitem[{Lam et~al.(2023)}]{Lam2023}
Lam, R., and Coauthors, 2023: Learning skillful medium-range global weather forecasting. \textit{Science}, \textbf{382~(6677)}, 1416--1421, \doi{10.1126/science.adi2336}.

\bibitem[{Larvor et~al.(2020)Larvor, Berthomier, Chabot, Le~Pape, Pradel,, and Perez}]{Meteonet}
Larvor, G., L.~Berthomier, V.~Chabot, B.~Le~Pape, B.~Pradel, and L.~Perez, 2020: Meteonet, an open reference weather dataset by meteo france. \url{https://github.com/meteofrance/meteonet}.

\bibitem[{Leinonen et~al.(2022)Leinonen, Hamann,, and Germann}]{leinonen2022}
Leinonen, J., U.~Hamann, and U.~Germann, 2022: Seamless {Lightning} {Nowcasting} with {Recurrent}-{Convolutional} {Deep} {Learning}. \textit{Artificial Intelligence for the Earth Systems}, \textbf{1~(4)}, e220\,043, \doi{10.1175/AIES-D-22-0043.1}.

\bibitem[{Liu et~al.(2023)}]{liu2023}
Liu, Q., and Coauthors, 2023: Deep-learning post-processing of short-term station precipitation based on {NWP} forecasts. \textit{Atmos.\ Res.}, \textbf{295}, 107\,032, \doi{10.1016/j.atmosres.2023.107032}.

\bibitem[{Mandement and Kreitz(2025)Mandement, and Kreitz}]{Valence}
Mandement, M., and M.~Kreitz, 2025: Précipitations et inondations exceptionnelles autour de valence. \textit{La Météorologie}, \textbf{~(128)}, \doi{10.37053/lameteorologie-2025-0003}.

\bibitem[{Mason(1979)}]{Mason1979}
Mason, I., 1979: On reducing probability forecasts to yes/no forecasts. \textit{Mon.\ Wea.\ Rev.}, \textbf{107~(2)}, 207 -- 211, \doi{10.1175/1520-0493(1979)107<0207:ORPFTY>2.0.CO;2}.

\bibitem[{McGovern et~al.(2019)McGovern, Lagerquist, John~Gagne, Jergensen, Elmore, Homeyer,, and Smith}]{Mcgovern2019}
McGovern, A., R.~Lagerquist, D.~John~Gagne, G.~E. Jergensen, K.~L. Elmore, C.~R. Homeyer, and T.~Smith, 2019: Making the {Black} {Box} {More} {Transparent}: {Understanding} the {Physical} {Implications} of {Machine} {Learning}. \textit{Bull.\ Amer.\ Meteor.\ Soc.}, \textbf{100~(11)}, 2175--2199, \doi{10.1175/BAMS-D-18-0195.1}.

\bibitem[{Murphy(2022)}]{Murphy2022}
Murphy, K.~P., 2022: \textit{Probabilistic machine learning: an introduction}. Adaptive computation and machine learning series, The MIT Press, Cambridge, Massachusetts.

\bibitem[{Nuissier et~al.(2011)Nuissier, Joly, Joly, Ducrocq,, and Arbogast}]{Nuissier2011}
Nuissier, O., B.~Joly, A.~Joly, V.~Ducrocq, and P.~Arbogast, 2011: A statistical downscaling to identify the large‐scale circulation patterns associated with heavy precipitation events over southern {France}. \textit{Quart.\ J.\ Roy.\ Meteor.\ Soc.}, \textbf{137~(660)}, 1812--1827, \doi{10.1002/qj.866}.

\bibitem[{Pardé(1941)}]{Aiguat}
Pardé, M., 1941: La formidable crue d'octobre 1940 dans les {Pyrénées}-{Orientales}. \textit{Revue géographique des Pyrénées et du Sud-Ouest. Sud-Ouest Européen}, \textbf{12~(3)}, 237--279, \doi{10.3406/rgpso.1941.4493}.

\bibitem[{{Peirce}(1884)}]{PSS}
{Peirce}, C.~S., 1884: {The Numerical Measure of the Success of Predictions}. \textit{Science}, \textbf{4~(93)}, 453--454, \doi{10.1126/science.ns-4.93.453}.

\bibitem[{Pirone et~al.(2023)Pirone, Cimorelli, Del~Giudice,, and Pianese}]{pirone2023}
Pirone, D., L.~Cimorelli, G.~Del~Giudice, and D.~Pianese, 2023: Short-term rainfall forecasting using cumulative precipitation fields from station data: a probabilistic machine learning approach. \textit{Journal of Hydrology}, \textbf{617}, 128\,949, \doi{10.1016/j.jhydrol.2022.128949}.

\bibitem[{Rasp and Lerch(2018)Rasp, and Lerch}]{rasp2018}
Rasp, S., and S.~Lerch, 2018: Neural {Networks} for {Postprocessing} {Ensemble} {Weather} {Forecasts}. \textit{Mon.\ Wea.\ Rev.}, \textbf{146~(11)}, 3885--3900, \doi{10.1175/MWR-D-18-0187.1}.

\bibitem[{Ricard et~al.(2012)Ricard, Ducrocq,, and Auger}]{Ricard2012}
Ricard, D., V.~Ducrocq, and L.~Auger, 2012: A {Climatology} of the {Mesoscale} {Environment} {Associated} with {Heavily} {Precipitating} {Events} over a {Northwestern} {Mediterranean} {Area}. \textit{J.\ Appl.\ Meteor.\ Climatol.}, \textbf{51~(3)}, 468--488, \doi{10.1175/JAMC-D-11-017.1}.

\bibitem[{Rodwell(2011)}]{Rodwell2011}
Rodwell, M.~J., 2011: On {Peirce}’s {Motivation} for {Equitability} in {Forecast} {Verification}. \textit{Mon.\ Wea.\ Rev.}, \textbf{139~(11)}, 3667--3669, \doi{10.1175/MWR-D-11-00167.1}.

\bibitem[{Scheffknecht et~al.(2016)Scheffknecht, Richard,, and Lambert}]{Scheffknecht2016}
Scheffknecht, P., E.~Richard, and D.~Lambert, 2016: A highly localized high-precipitation event over corsica. \textit{Quart.\ J.\ Roy.\ Meteor.\ Soc.}, \textbf{142~(S1)}, 206--221, \doi{https://doi.org/10.1002/qj.2795}, \eprint{https://rmets.onlinelibrary.wiley.com/doi/pdf/10.1002/qj.2795}.

\bibitem[{Seity et~al.(2011)Seity, Brousseau, Malardel, Hello, Bénard, Bouttier, Lac,, and Masson}]{Arome}
Seity, Y., P.~Brousseau, S.~Malardel, G.~Hello, P.~Bénard, F.~Bouttier, C.~Lac, and V.~Masson, 2011: The {AROME}-{France} {Convective}-{Scale} {Operational} {Model}. \textit{Mon.\ Wea.\ Rev.}, \textbf{139~(3)}, \doi{10.1175/2010MWR3425.1}.

\bibitem[{Shi et~al.(2015)Shi, Chen, Wang, Yeung, Wong,, and Woo}]{shi2015}
Shi, X., Z.~Chen, H.~Wang, D.-Y. Yeung, W.-k. Wong, and W.-c. Woo, 2015: Convolutional {LSTM} {Network}: {A} {Machine} {Learning} {Approach} for {Precipitation} {Nowcasting}.

\bibitem[{Shi et~al.(2017)Shi, Gao, Lausen, Wang, Yeung, Wong,, and Woo}]{shi2017}
Shi, X., Z.~Gao, L.~Lausen, H.~Wang, D.-Y. Yeung, W.-k. Wong, and W.-c. Woo, 2017: Deep {Learning} for {Precipitation} {Nowcasting}: {A} {Benchmark} and {A} {New} {Model}. arXiv, \urlprefix\url{http://arxiv.org/abs/1706.03458}, arXiv:1706.03458 [cs].

\bibitem[{Wang et~al.(2017)}]{WMO2017}
Wang, Y., and Coauthors, 2017: \textit{Guidelines for Nowcasting Techniques}.

\bibitem[{Warner(2010)}]{Warner2010}
Warner, T.~T., 2010: \textit{Numerical Weather and Climate Prediction}. Cambridge University Press.

\bibitem[{Wilks(2019)}]{WILKS2019369}
Wilks, D.~S., 2019: Chapter 9 - forecast verification. \textit{Statistical Methods in the Atmospheric Sciences (Fourth Edition)}, D.~S. Wilks, Ed., fourth edition ed., Elsevier, 369--483, \doi{https://doi.org/10.1016/B978-0-12-815823-4.00009-2}, \urlprefix\url{https://www.sciencedirect.com/science/article/pii/B9780128158234000092}.

\bibitem[{Woodcock(1976)}]{Woodcock1976}
Woodcock, F., 1976: The evaluation of yes/no forecasts for scientific and administrative purposes. \textit{Mon.\ Wea.\ Rev.}, \textbf{104~(10)}, 1209 -- 1214, \doi{10.1175/1520-0493(1976)104<1209:TEOYFF>2.0.CO;2}.

\end{thebibliography}
\end{document}